\documentclass[journal]{IEEEtran}

\ifCLASSINFOpdf
\else
\fi
\usepackage[dvips]{graphicx}
\usepackage{latexsym}
\usepackage{amssymb}
\usepackage{amsmath}
\usepackage{bm}
\usepackage{multirow}
\usepackage{xcolor}
\usepackage{lipsum}
\usepackage{multicol}
\usepackage{enumerate}
\usepackage{algorithm}
\usepackage{algorithmicx}
\usepackage{algpseudocode}
\usepackage{amsmath}
\usepackage{graphicx}
\usepackage{subfig}
\usepackage[utf8]{inputenc}
\usepackage{float}

\begin{document}
%
\title{Interference Exploitation 1-Bit Massive MIMO Precoding: A Partial Branch-and-Bound Solution with Near-Optimal Performance}
%
%
%

\author{Ang~Li,~\IEEEmembership{Member,~IEEE}, Fan Liu,~\IEEEmembership{Member,~IEEE}, Christos Masouros,~\IEEEmembership{Senior Member,~IEEE},\\ Yonghui Li,~\IEEEmembership{Fellow,~IEEE}, and Branka Vucetic,~\IEEEmembership{Fellow,~IEEE}

\thanks{Manuscript received XX; revised XX; }
\thanks{A. Li, Y. Li and B. Vucetic are with the School of Electrical and Information Engineering, University of Sydney, Sydney, NSW 2006, Australia. (e-mail: \{ang.li2, yonghui.li, branka.vucetic\}@sydney.edu.au)}
\thanks{F. Liu and C. Masouros are with the Department of Electronic and Electrical Engineering, University College London, Torrington Place, London, WC1E 7JE, UK (e-mail: \{fan.liu, c.masouros\}@ucl.ac.uk).}
\thanks{This work was supported in part by the European Union's Horizon 2020 research and innovation programme under the Marie Sk\l{}odowska-Curie Grant Agreement No. 793345, in part by the EPSRC project EP/R007934/1, in part by the ...}
}



\maketitle

\begin{abstract}
In this paper, we focus on 1-bit precoding approaches for downlink massive multiple-input multiple-output (MIMO) systems, where we exploit the concept of constructive interference (CI). For both PSK and QAM signaling, we firstly formulate the optimization problem that maximizes the CI effect subject to the requirement of the 1-bit transmit signals. We then mathematically prove that, when employing the CI formulation and relaxing the 1-bit constraint, the majority of the transmit signals already satisfy the 1-bit formulation. Building upon this important observation, we propose a 1-bit precoding approach that further improves the performance of the conventional 1-bit CI precoding via a partial branch-and-bound (P-BB) process, where the BB procedure is performed only for the entries that do not comply with the 1-bit requirement. This operation allows a significant complexity reduction compared to the fully-BB (F-BB) process, and enables the BB framework to be applicable to the complex massive MIMO scenarios. We further develop an alternative 1-bit scheme through an `Ordered Partial Sequential Update' (OPSU) process that allows an additional complexity reduction. Numerical results show that both proposed 1-bit precoding methods exhibit a significant signal-to-noise ratio (SNR) gain for the error rate performance, especially for higher-order modulations.

\end{abstract}


%
\IEEEpeerreviewmaketitle

\section{Introduction}
%
%
%
%

\IEEEPARstart{M}{ASSIVE} multiple-input multiple-output (MIMO) has become a key enabling technology for the fifth-generation (5G) and future wireless communication systems \cite{intro1}\nocite{intro2}\nocite{intro3}\nocite{intro4}\nocite{intro5}-\cite{intro6}. In the downlink transmission of a massive MIMO system, existing non-linear precoding methods such as Tomlinson-Harashima precoding (THP) \cite{THP} or vector perturbation (VP) precoding \cite{VP1}\nocite{VP2}\nocite{VP3}-\cite{VP4} are not preferred, due to their prohibitive computational complexity when the number of antennas is large. Instead, it has been shown in \cite{intro7} that low-complexity linear precoding approaches such as zero-forcing (ZF) \cite{ZF} and regularized ZF (RZF) \cite{RZF} can achieve near-optimal performance.

The near optimality for linear precoding in massive MIMO is achieved assuming that fully-digital processing and high-resolution digital-to-analog converters (DACs) are employed at the base station (BS). However, this fully-digital processing requires a dedicated radio frequency (RF) chain and a pair of high-resolution DACs for each antenna element, which results in a significant increase in the hardware complexity and cost when the number of transmit antennas scales up. Moreover, the resulting power consumption of the large number of hardware components will also be prohibitive for practical implementation. All of the above drawbacks make fully-digital processing highly undesirable for a massive MIMO BS. Accordingly, there have been several emerging techniques that aim to reduce the hardware complexity and the power consumption for a massive MIMO BS, including hybrid analog-digital (AD) precoding \cite{hybrid1}\nocite{hybrid2}\nocite{hybrid3}\nocite{hybrid4}\nocite{hybrid5}\nocite{hybrid6}\nocite{hybrid7}-\cite{hybrid8}, constant-envelope (CE) precoding \cite{cep1}\nocite{cep2}\nocite{cep3}-\cite{cep4}, and low-resolution DACs.

Hybrid AD precoding reduces the hardware complexity and cost by reducing the number of RF chains, where precoding is divided into the analog domain and the low-dimension fully-digital domain \cite{hybrid1}. CE precoding reduces the hardware complexity by transmitting CE signals, which allows the use of the most power-efficient and cheapest RF amplifiers for each RF chain \cite{cep3}. In addition to the above two techniques, the use of low-resolution DACs, which is the focus of this paper, can reduce the hardware cost and power consumption per RF chain by reducing the resolution of the DACs. Since the power consumption of DACs grows exponentially with the resolution and linearly with the bandwidth \cite{dac2}, \cite{dac3}, adopting low-resolution DACs instead of high-resolution ones can greatly reduce the power consumption at the BS, especially in the case of massive MIMO where a large number of DACs are required. Among low-resolution DACs, the most extreme case, i.e., 1-bit DACs, has received particular research interest, not only because it allows the most significant power savings, but also because the output signals of 1-bit DACs are CE signals, which further enables the use of the most power-efficient RF amplifiers, as in the case for CE precoding.

In the existing literature, there have already been some works that consider the precoding designs in the presence of 1-bit DACs \cite{dac1}\nocite{dac4}-\cite{dac5}. In \cite{dac1}, the traditional ZF precoding was applied to the case of 1-bit DACs, where the 1-bit quantization was directly performed upon the ZF precoded signals, and an error floor is observed as the transmit signal-to-noise ratio (SNR) increases. The significant performance loss is as expected for this naive precoding method. In \cite{dac4}, a 1-bit quantized linear precoding method was proposed based on the minimum mean squared error (MMSE) metric, which achieves an improved performance over the quantized ZF precoding approach. In \cite{dac5}, the 1-bit precoding algorithm was proposed via an iterative gradient projection process based on the MMSE metric. However, error floors can still be observed for the 1-bit precoding schemes proposed in \cite{dac4} and \cite{dac5}, which result from the fact that linear precoding is still considered, i.e., the precoded signals before quantization are linear transformations of the data symbols. To further improve the error rate performance, non-linear 1-bit precoding designs, which directly map the data symbols into the 1-bit transmit signals through a symbol-level operation, were further proposed in \cite{dac6}\nocite{dac7}\nocite{dac8}\nocite{dac9}\nocite{dac10}\nocite{dac11}\nocite{dac12}\nocite{dac13}\nocite{dac14}-\cite{dac15}. In \cite{dac6} and \cite{dac7}, non-linear 1-bit precoding schemes were proposed via the gradient projection algorithm based on the minimum bit error rate (BER) metric and MMSE metric, respectively. Both proposed 1-bit algorithms outperform \cite{dac1}-\cite{dac5} significantly, especially in medium-to-high SNR regime. \cite{dac8} proposed a 1-bit precoding design via a biconvex relaxation procedure, while \cite{dac9} extended the work in \cite{dac8} and proposed several 1-bit precoding schemes based on semidefinite relaxation (SDR), $\ell_\infty$-norm relaxation, and sphere precoding, respectively. \cite{dac11} improves the performance of the schemes proposed in \cite{dac9} through an alternating optimization framework, when a high-order QAM modulation is adopted at the BS. 

Nevertheless, it should be noted that these MMSE-based precoding methods may be sub-optimal since they ignore that multi-user interference can be constructive and further benefit the performance, when symbol-level precoding is employed. Considering a PSK constellation as an example, if the received signal is forced to locate deeper within the decision region and further away from the detection boundaries, a more reliable decoding performance can be obtained, though the MSE in this case will increase. This observation has already been exploited in \cite{VP3} and \cite{ci1}\nocite{ci2}\nocite{ci3}-\cite{ci14} by constructive interference (CI) precoding to achieve an improved BER performance in a traditional small-scale MIMO system. Following this concept, \cite{dac12} and \cite{dac13} have extended the idea of interference exploitation to 1-bit precoding designs, and the resulting BER performance is shown to be promising. Moreover, while not explicitly shown, \cite{dac14} also adopts the formulation of CI-based 1-bit precoding, where a branch-and-bound (BB)-based algorithm that obtains the optimal solution is presented. More recently, the BB framework has been extended to the case of QAM modulations in \cite{dac15} based on the QR decomposition. However, the above two 1-bit designs based on the fully-BB (F-BB) process are still not practically useful in massive MIMO systems due to their unfavorable complexity.

In this paper, we focus on designing a near-optimal 1-bit precoding algorithm as well as its low-complexity variation for massive MIMO systems, where both PSK and QAM modulations are considered. We exploit the concept of CI to formulate the optimization problem, which aims to maximize the CI effect subject to the 1-bit output signal requirement. The proposed near-optimal 1-bit precoding solution is achieved via a judicious partial BB (P-BB) procedure, while its low-complexity counterpart is implemented through a greedy algorithm. For clarity, we summarize the main contributions of this paper below:

\begin{enumerate}
\item For both PSK and QAM signaling, by constructing the Lagrangian function of the relaxed optimization problem and formulating the corresponding Karush-Kuhn-Tucker (KKT) conditions, we mathematically prove by contradiction that the majority of the output signals obtained from solving the relaxed problem already satisfy the 1-bit constraint, and only a small portion of the entries need to be further quantized to obtain a feasible 1-bit solution, where the quantization losses are incurred.

\item Building on this important and interesting observation, we propose a 1-bit precoding algorithm through a P-BB process to further improve the performance of the conventional CI-based 1-bit precoding method in \cite{dac13}, where the BB process is only performed for part of the output signals that do not comply with the 1-bit requirement, and we adopt the adaptive subdivision rule to guarantee a faster convergence rate. For PSK signaling, we use the `max-min' criterion to design the P-BB algorithm, while the MSE criterion and the alternating optimization framework are employed when QAM signaling is considered at the BS. Compared to the conventional F-BB method whose complexity becomes prohibitive in massive MIMO scenarios, our proposed P-BB approach enables the use of the BB framework in massive MIMO systems and allows a significant gain in terms of computational cost, while still exhibiting a near-optimal error rate performance. 

\item We further design an alternative 1-bit precoding scheme through an `Ordered Partial Sequential Update' (OPSU) process, where we only consider the effect of a single entry at a time on the objective function, while keeping other entries in the output signals fixed. The proposed OPSU method further allows an additional complexity reduction compared to the P-BB approach, and is particularly appealing when the P-BB process needs to search the entire subspace. 

\item Compared to the conventional CI-based approach and other existing 1-bit precoding methods in the literature, numerical results demonstrate an SNR gain of more than 7dB for the proposed 1-bit precoding schemes in terms of BER, which also remove the error floors that are commonly observed in conventional 1-bit precoding techniques, especially when higher-order modulations are adopted at the BS.

\end{enumerate}

The remainder of this paper is organized as follows. Section II introduces the basic system model and concept of CI. Section III includes the proposed 1-bit precoding approaches for PSK signaling, and Section IV extends the proposed 1-bit precoding schemes to QAM signaling. Numerical results are shown in Section V, and Section VI concludes our paper.

{\bf Notations:} $a$, $\bf a$, and $\bf A$ denote scalar, column vector and matrix, respectively. ${( \cdot )^\text{T}}$ and ${( \cdot )^\text{H}}$ denote transposition and conjugate transposition of a matrix, respectively. $\text{card} \left( {\cdot} \right)$ denotes the cardinality of a set, ${\text {sgn}} \left[  \cdot  \right]$ is the sign function, and $\jmath$ denotes the imaginary unit. $\left|  \cdot  \right|$ denotes the modulus of a complex number or the absolute value of a real number, and $\left\|  \cdot  \right\|_2$ denotes the $\ell_2$-norm. ${{\cal C}^{n \times n}}$ and ${{\cal R}^{n \times n}}$ represent an $n \times n$ matrix in the complex  and real set, respectively. $\Re ( \cdot )$ and $\Im ( \cdot )$ denote the real and imaginary part of a complex number, respectively. ${\text {rank}} \left( \cdot \right)$ returns the rank of a matrix, and ${\bf I}_K$ represents a $K \times K$ identity matrix.

\begin{figure}[!t]
\centering
\includegraphics[scale=0.45]{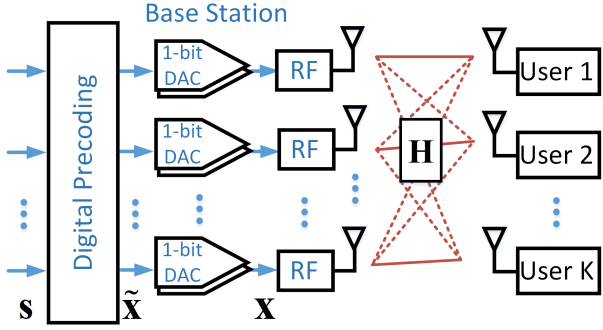}
\caption{A downlink massive MIMO system with 1-bit DACs}
\end{figure}

\section{System Model and Constructive Interference}
\subsection{System Model}
We consider a massive MIMO system in the downlink, as depicted in Fig.~1, where a BS with $N_t$ transmit antennas communicates with a total number of $K$ single-antenna users simultaneously in the same time-frequency resource, where $K \ll N_t$. As we focus on the precoding design at the BS, ideal ADCs are employed for each user, and we assume perfect knowledge of CSI is known \cite{dac4}-\cite{dac11}. We denote the data symbol vector as ${\bf s} \in {\cal C}^{K \times 1}$, which can be drawn from a unit-norm PSK or a normalized QAM constellation. We denote ${\bf H}\in {\cal C}^{K \times N_t}$ as the flat-fading Rayleigh channel matrix between the BS and the users, with each entry following a standard complex Gaussian distribution $\mathbb {CN}\left( {0,1} \right)$. The corresponding transmit signal vector before quantization can then be expressed as
\begin{equation}
{\bf \tilde x} = {\cal P}\left( {{\bf{s}},{\bf{H}}} \right),
\label{eq_1}
\end{equation}
which is a function of the symbol vector $\bf s$ as well as the channel matrix $\bf H$. $\cal P$ represents a general precoding strategy that forms the desired unquantized signal vector $\bf \tilde x$, which can be a linear transformation of $\bf s$ as in \cite{dac4}-\cite{dac6} or a non-linear mapping as in \cite{dac7}-\cite{dac15}. When 1-bit DACs are adopted at the BS, the output signal vector on the antenna elements is given by
\begin{equation}
{\bf x}={\cal Q}\left( {\bf \tilde x} \right),
\label{eq_2}
\end{equation}
where $\cal Q$ is the element-wise 1-bit quantization on both real and imaginary part of $\bf \tilde x$. For simplicity, we normalize $\bf x$ such that $\left\| {\bf{x}} \right\|_2^2 = 1$, which leads to
\begin{equation}
x_n \in {\cal X}_{{\text{DAC}}}, {\kern 3pt} \forall n \in {\cal N},
\label{eq_3}
\end{equation}
where $x_n$ is the $n$-th entry in $\bf x$, ${\cal X}_{\text {DAC}}=\left\{ {\pm \frac{1}{{\sqrt {2{N_t}} }} \pm \frac{1}{{\sqrt {2{N_t}} }}\jmath} \right\}$, and ${\cal N}=\left\{ {1,2,\cdots,N_t} \right\}$. Accordingly, the received signal vector ${\bf y} \in {\cal C}^{K \times 1}$ can be expressed as
\begin{equation}
{\bf y}= {\bf Hx} + {\bf n},
\label{eq_4}
\end{equation}
where ${\bf n} \in {\cal C}^{K \times 1}$ is the additive Gaussian noise at the receiver side and ${\bf n} \sim {\mathbb {CN}}\left( {0,{\sigma ^2} \cdot {\bf{I}}_K} \right)$.

\subsection{Constructive Interference}
CI is defined as the interference that leads to an increased distance to all the detection thresholds for a specific constellation point, as discussed in \cite{ci1}-\cite{ci3}. Closed-form CI precoding was firstly considered for PSK signaling in small-scale MIMO systems to improve the performance of the linear ZF precoding in \cite{ci4}\nocite{ci5}-\cite{ci6}. The optimization-based CI approach firstly appeared in \cite{VP3}, and has more recently been extensively studied in \cite{ci7}\nocite{ci8}\nocite{ci9}\nocite{ci10}-\cite{ci11}, where the constructive area is introduced. It is shown that, as long as the received signal is located within the constructive area, the corresponding interfering signals are beneficial, which further improve the error rate performance. CI precoding has further been extended to QAM constellations in \cite{ci12}, \cite{ci13}. Compared to PSK modulations where all the constellation points can exploit CI, only part of the constellation points for QAM modulations can exploit CI, since we observe all the interference for the inner constellation points of QAM to be destructive, as discussed in \cite{ci13}.

\begin{figure}[!b]
\centering
\includegraphics[scale=0.4]{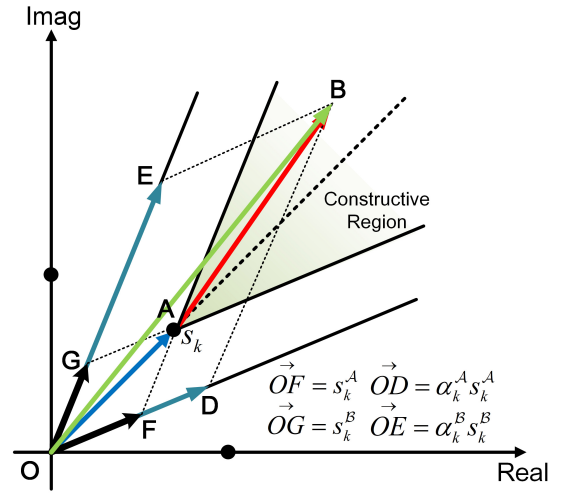}
\caption{An illustrative example of CI condition for PSK}
\end{figure}

\section{1-Bit Precoding for PSK Signaling}
\subsection{CI Condition and Problem Formulation}
Before presenting the 1-bit precoding designs, we first briefly introduce the mathematical formulation of the CI condition for PSK modulations based on the `symbol-scaling' metric, as depicted in Fig.~2, where we adopt one quarter of an 8PSK constellation as the example \cite{ci13}. Without loss of generality, we express
\begin{equation}
\vec{OA}=s_k=s_k^{\cal A} + s_k^{\cal B}
\label{eq_5}
\end{equation}
to denote a unit-norm constellation point, where we have further decomposed the constellation point into $\vec{OF}=s_k^{\cal A}$ and $\vec{OG}=s_k^{\cal B}$ that are parallel to the two detection boundaries of $s_k$. The detailed expressions for $s_k^{\cal A}$ and $s_k^{\cal B}$ can be found in the appendix of \cite{dac13} for a general $\mathbb M$-PSK modulation, and are omitted here for brevity. $\vec{OB}=\vec{OD}+\vec{OE}$ denotes the received signal for user $k$ excluding noise, which is similarly decomposed into
\begin{equation}
\vec{OB}={\bf h}_k^\text{T} {\bf x}= \alpha_k^{\cal A} s_k^{\cal A} + \alpha_k^{\cal B} s_k^{\cal B},
\label{eq_6}
\end{equation}
where ${\bf h}_k^\text{T}$ is the $k$-th row of $\bf H$. $\alpha_k^{\cal A}$ and $\alpha_k^{\cal B}$ are two introduced real auxiliary variables that fully represent the effect of interference and 1-bit quantization on $s_k$. Following \cite{VP3} and \cite{ci13}, the `symbol-scaling' CI condition for PSK signaling can be expressed as
\begin{equation}
\alpha _k^{\cal U} \ge 0, {\kern 3pt} \forall k \in {\cal K}, {\kern 3pt} {\cal U} \in \left\{ {{\cal A}, {\cal B}} \right\},
\label{eq_7}
\end{equation}
where ${\cal K}=\left\{ {1,2,\cdots,K} \right\}$. Accordingly, the 1-bit precoding design that exploits CI and maximizes its effect can be formulated as
\begin{equation}
\begin{aligned}
&\mathcal{P}_1: {\kern 3pt} \mathop {\max }\limits_{\bf{x}} \mathop {\min }\limits_{k, {\kern 1pt} {\cal U}} {\kern 3pt} \alpha _k^{\cal U} \\
&{\kern 2pt} \text{s.t.} {\kern 10pt} {{\bf{h}}_k^\text{T}}{\bf x} = \alpha_k^{\cal A} s_k^{\cal A} + \alpha_k^{\cal B} s_k^{\cal B}, {\kern 3pt} \forall k \in {\cal K} \\
&{\kern 24pt} x_n \in {\cal X}_{{\text{DAC}}}, {\kern 3pt} \forall n \in {\cal N} \\
&{\kern 24pt} {\cal U} \in \left\{ {{\cal A}, {\cal B}} \right\}
\label{eq_8}
\end{aligned}
\end{equation}

${\cal P}_1$ is a non-convex optimization problem due to the 1-bit constraint $x_n \in {\cal X}_{{\text{DAC}}}$, $\forall n \in {\cal N}$, and it is therefore difficult to directly obtain the optimal solution. Nevertheless, by relaxing this non-convex constraint, ${\cal P}_1$ can readily be transformed into a convex problem:
\begin{equation}
\begin{aligned}
&\mathcal{P}_2: {\kern 3pt} \mathop {\max }\limits_{\bf \tilde x} \mathop {\min }\limits_{k, {\kern 1pt} {\cal U}} {\kern 3pt} \alpha _k^{\cal U} \\
&{\kern 2pt} \text{s.t.} {\kern 10pt} {{\bf{h}}_k^\text{T}}{\bf \tilde x} = \alpha_k^{\cal A} s_k^{\cal A} + \alpha_k^{\cal B} s_k^{\cal B}, {\kern 3pt} \forall k \in {\cal K} \\
&{\kern 22pt} \left| {\Re \left( {{{\tilde x}_n}} \right)} \right| \le \frac{1}{{\sqrt {2{N_t}} }}, {\kern 3pt} \forall n \in {\cal N}\\
&{\kern 22pt} \left| {\Im \left( {{{\tilde x}_n}} \right)} \right| \le \frac{1}{{\sqrt {2{N_t}} }}, {\kern 3pt} \forall n \in {\cal N}\\
&{\kern 24pt} {\cal U} \in \left\{ {{\cal A}, {\cal B}} \right\}
\label{eq_9}
\end{aligned}
\end{equation}
where $\tilde x_n$ is the $n$-th entry in $\bf \tilde x$. With the relaxed signal vector $\bf \tilde x$ obtained by solving ${\cal P}_2$, a feasible solution to the original 1-bit precoding problem ${\cal P}_1$ can be obtained by enforcing an element-wise normalization, given by
\begin{equation}
{x_n} = \frac{{{\text {sgn}} \left[ {\Re \left( {{{\tilde x}_n}} \right)} \right]}}{{\sqrt {2{N_t}} }} + \frac{{{\text {sgn}} \left[ {\Im \left( {{{\tilde x}_n}} \right)} \right]}}{{\sqrt {2{N_t}} }}\jmath, {\kern 3pt} \forall n \in {\cal N}.
\label{eq_10}
\end{equation}
For notational simplicity, we denote the final quantized signal vector and the 1-bit precoding scheme based on the above relaxation-normalization procedure as ${\bf{x}}_\text{CI}^\text{PSK}$ and `CI 1-Bit', respectively.

\subsection{Analytical Study of 1-Bit CI Precoding for PSK}
It has been shown in \cite{dac13} that the error rate performance of `CI 1-Bit' is promising, which outperforms many of the existing 1-bit precoding designs in the literature for PSK signaling \cite{dac4}-\cite{dac8}. In fact, it is numerically observed in \cite{dac13} that most of the entries in $\bf \tilde x$ obtained by solving ${\cal P}_2$ already satisfy the 1-bit constraint, while an element-wise relaxation is performed afterwards. This is the main reason why the performance of `CI 1-Bit' is promising, since only a small part of the entries in $\bf \tilde x$ need to be further quantized, which leads to an insignificant quantization loss. Nevertheless, \cite{dac13} fails to explain this observation from a mathematical point of view.

In this section, we further elaborate on this observation, and propose a 1-bit precoding method via the P-BB method based on this observation, which further improves the performance of `CI 1-Bit' and achieves a close-to-optimal error rate performance. To begin with, we first transform the relaxed optimization problem ${\cal P}_2$ into a simpler form for ease of our analysis. By comparing the real and imaginary part of both sides of \eqref{eq_6}, we can express $\alpha _k^{\cal U}$ as a function of $\bf H$ and $\bf s$, given by
\begin{equation}
\begin{aligned}
\alpha _k^{\cal A} =& {\kern 2pt} \frac{{\Im \left( {s_k^{\cal B}} \right)\Re \left( {{\bf{h}}_k^\text{T}} \right) - \Re \left( {s_k^{\cal B}} \right)\Im \left( {{\bf{h}}_k^\text{T}} \right)}}{{\Re \left( {s_k^{\cal A}} \right)\Im \left( {s_k^{\cal B}} \right) - \Im \left( {s_k^{\cal A}} \right)\Re \left( {s_k^{\cal B}} \right)}} \cdot \Re \left( {\bf{x}} \right)\\
& - \frac{{\Im \left( {s_k^{\cal B}} \right)\Im \left( {{\bf{h}}_k^\text{T}} \right) + \Re \left( {s_k^{\cal B}} \right)\Re \left( {{\bf{h}}_k^\text{T}} \right)}}{{\Re \left( {s_k^A} \right)\Im \left( {s_k^B} \right) - \Im \left( {s_k^{\cal A}} \right)\Re \left( {s_k^{\cal B}} \right)}} \cdot \Im \left( {\bf{x}} \right) \\
=& {\kern 2pt} {\bf{a}}_k^\text{T}\Re \left( {\bf{x}} \right) + {\bf{b}}_k^\text{T}\Im \left( {\bf{x}} \right), \\
\alpha _k^{\cal B} =& {\kern 2pt} \frac{{\Re \left( {s_k^{\cal A}} \right)\Im \left( {{\bf{h}}_k^\text{T}} \right) - \Im \left( {s_k^{\cal A}} \right)\Re \left( {{\bf{h}}_k^\text{T}} \right)}}{{\Re \left( {s_k^{\cal A}} \right)\Im \left( {s_k^{\cal B}} \right) - \Im \left( {s_k^{\cal A}} \right)\Re \left( {s_k^{\cal B}} \right)}} \cdot \Re \left( {\bf{x}} \right)\\
& + \frac{{\Re \left( {s_k^{\cal A}} \right)\Re \left( {{\bf{h}}_k^\text{T}} \right) + \Im \left( {s_k^{\cal A}} \right)\Im \left( {{\bf{h}}_k^\text{T}} \right)}}{{\Re \left( {s_k^{\cal A}} \right)\Im \left( {s_k^{\cal B}} \right) - \Im \left( {s_k^{\cal A}} \right)\Re \left( {s_k^{\cal B}} \right)}} \cdot \Im \left( {\bf{x}} \right) \\
=& {\kern 2pt} {\bf{c}}_k^\text{T}\Re \left( {\bf{x}} \right) + {\bf{d}}_k^\text{T}\Im \left( {\bf{x}} \right).
\end{aligned}
\label{eq_11}
\end{equation}
By defining
\begin{equation}
{\bf{p}}_k^\text{T} = \left[ {{\bf{a}}_k^\text{T},{\bf{b}}_k^\text{T}} \right], {\kern 2pt} {\bf{q}}_k^\text{T} = \left[ {{\bf{c}}_k^\text{T},{\bf{d}}_k^\text{T}} \right], {\kern 2pt} {{\bf{x}}_{\bf{E}}} = {\left[ {\Re \left( {{{\bf{x}}^\text{T}}} \right),\Im \left( {{{\bf{x}}^\text{T}}} \right)} \right]^\text{T}},
\label{eq_12}
\end{equation}
and
\begin{equation}
{\bf \Lambda} = {\left[ {\alpha _1^{\cal A}, \alpha _2^{\cal A}, \cdots ,\alpha _K^{\cal A},\alpha _1^{\cal B}, \alpha _2^{\cal B}, \cdots ,\alpha _K^{\cal B}} \right]^\text{T}}, 
\label{eq_13}
\end{equation}
\eqref{eq_11} can be expressed in a compact matrix form as
\begin{equation}
{\bf \Lambda}= {\bf M}{\bf x_E},
\label{eq_14}
\end{equation}
where ${\bf M} \in {\cal R}^{2K \times 2N_t}$ is given by
\begin{equation}
{\bf{M}} = {\left[ {{\bf{p}}_1,{\bf{p}}_2, \cdots ,{\bf{p}}_K,{\bf{q}}_1,{\bf{q}}_2, \cdots ,{\bf{q}}_K} \right]^\text{T}}.
\label{eq_15}
\end{equation}
Based on the construction of $\bf M$ shown above, the following rank property is observed.

{\bf Lemma 1:} ${\text {rank}}\left( {\bf M} \right)=2K$ with probability 1.

{\bf Proof:} See Appendix A. ${\kern 130pt} \blacksquare$

With the matrix formulation in \eqref{eq_14}, the relaxed optimization problem ${\cal P}_2$ is equivalent to 
\begin{equation}
\begin{aligned}
&\mathcal{P}_3: {\kern 3pt} \mathop {\max }\limits_{{\bf \tilde x}_{\bf E}} \mathop {\min }\limits_{l} {\kern 3pt} \alpha _l \\
&{\kern 2pt} \text{s.t.} {\kern 10pt} \alpha _l= {\bf m}_l^\text{T}{\bf \tilde x}_{\bf E}, {\kern 3pt} \forall l \in {\cal L} \\
&{\kern 22pt} \left| {{{\tilde x}_m^{\text E}}} \right| \le \frac{1}{{\sqrt {2{N_t}} }}, {\kern 3pt} \forall m \in {\cal M}
\label{eq_22}
\end{aligned}
\end{equation}
where ${\bf m}_l^\text{T}$ is the $l$-th row of $\bf M$, ${{\tilde x}_n^{\text E}}$ is the $n$-th entry of ${\bf \tilde x}_{\bf E}$, ${\cal L}=\left\{ {1,2,\cdots,2K} \right\}$, and ${\cal M}=\left\{ {1,2,\cdots,2N_t} \right\}$.

Based on the formulation of ${\cal P}_3$, the following important proposition is obtained, which builds the foundation of the proposed 1-bit precoding algorithms through P-BB in the following.

{\bf Proposition 1:} For ${\bf \tilde x}_{\bf E}$ obtained by solving ${\cal P}_3$, there are at least $\left( {2{N_t} - 2K + 1} \right)$ entries that already satisfy the 1-bit constraint. 

{\bf Proof:} See Appendix B. ${\kern 130pt} \blacksquare$

{\bf Lemma 2:} The results of {\bf Proposition 1} directly extend to rank-deficient channels, where in this case there are at least $\left[ {2N_t-2 \cdot {\text {rank}} \left( {\bf H} \right)+1} \right]$ entries in $\bf \tilde x_E$ obtained by solving ${\cal P}_3$ that already satisfy the 1-bit constraint.

{\bf Proof:} The proof for this lemma follows the proof for {\bf Proposition 1}, and is therefore omitted for brevity. {\kern 27pt} $\blacksquare$

{\bf Proposition 1} mathematically explains the observation in \cite{dac13} and the reason why the performance of `CI 1-Bit' is promising. In the case of a massive MIMO system where $N_t \gg K$, $\left({2N_t-2K+1} \right)$ is close to $2N_t$, i.e., the majority of the entries in ${\bf \tilde x}_{\bf E}$ obtained by solving ${\cal P}_3$ already satisfy the 1-bit constraint, and the performance loss incurred from the subsequent quantization on the residual $\left( 2K-1 \right)$ (or even smaller) entries in ${\bf \tilde x}_{\bf E}$ becomes insignificant. Moreover, the performance loss to the optimal solution is expected to become even less for rank-deficient channels, where ${\text {rank}} \left( {\bf H} \right) < K$, as shown by {\bf Lemma 2}.

\subsection{1-Bit Precoding Design via Partial Branch-and-Bound}
Building upon the important observation in {\bf Proposition 1}, we introduce the 1-bit precoding method based on P-BB in this section. Essentially, as opposed to the F-BB method in \cite{dac14} that searches the entire space ${\cal X}_{\text {DAC}}^{2N_t}$, our proposed P-BB scheme only focuses on part of the space, i.e., ${\cal X}_{\text {DAC}}^{N_R}$, which corresponds to the entries in $\bf \tilde x_E$ that do not comply with the 1-bit constraint, where $N_R \le \left( {2K-1} \right)$. Therefore, compared to the F-BB scheme in \cite{dac14} whose complexity is proportional to the number of transmit antennas, which thus only works in small-scale MIMO systems, the P-BB approach introduced in this paper, whose complexity is only proportional to the number of users, enables a significant reduction in the computational cost of the BB-based method and allows the BB framework to be applicable in massive MIMO systems. 

To be more specific, we firstly conduct some row rearrangements for ${{\bf{\tilde x}}_{\bf{E}}}$ obtained from solving ${\cal P}_3$ to arrive at ${{\bf{\hat x}}_{\bf{E}}}$, such that ${{\bf{\hat x}}_{\bf{E}}}$ can be decomposed into
\begin{equation}
{{\bf{\hat x}}_{\bf{E}}} = {\left[ {{\bf{x}}_{\bf{F}}^\text{T},{\bf{x}}_{\bf{R}}^\text{T}} \right]^\text{T}},
\label{eq_31}
\end{equation}
where ${\bf{x}}_{\bf{F}} \in {\cal R}^{N_F \times 1}$ consists of $x_m^{\text E}$ that already satisfy the 1-bit constraint, and we obtain $N_F \ge \left({2N_t-2K+1}\right)$ following {\bf Proposition 1}. We further express ${\bf {x}}_{\bf{R}}=\left[ {x_1^{\text R},x_2^{\text R}, \cdots ,x_{N_R}^{\text R}} \right]^\text{T}$ that consists of the residual entries in ${{\bf{\hat x}}_{\bf{E}}}$ whose amplitudes are strictly smaller than $\frac{1}{\sqrt {2N_t}}$, where we have $N_R \le 2K-1$ and $N_F+N_R=2N_t$. We further denote the matrix $\bf M$ with the corresponding column rearrangement as $\bf \hat M$ (${\bf \hat M}{\bf \hat x_E}={\bf Mx_E}$), which is decomposed into
\begin{equation}
{\bf \hat{M}} = \left[ {{{\bf{M}}_{\bf{F}}},{{\bf{M}}_{\bf{R}}}} \right],
\label{eq_32}
\end{equation}
where ${{\bf{M}}_{\bf{F}}} = {\left[ {{\bf \hat{m}}_1^\text{F},{\bf \hat{m}}_2^\text{F}, \cdots ,{\bf \hat{m}}_{2K}^\text{F}} \right]^\text{T}} \in {\cal R}^{2K\times N_F}$ and ${{\bf{M}}_{\bf{R}}} = {\left[ {{\bf \hat{m}}_1^\text{R},{\bf \hat{m}}_2^\text{R}, \cdots ,{\bf \hat{m}}_{2K}^\text{R}} \right]^\text{T}} \in {\cal R}^{2K\times N_R}$. The resulting optimization problem on ${\bf {x}}_{\bf{R}}$ is then given by
\begin{equation}
\begin{aligned}
&\mathcal{P}_4: {\kern 3pt} \mathop {\min }\limits_{{{\bf{x}}_{\bf{R}}}} {\kern 3pt} - t \\
&{\kern 2pt} \text{s.t.} {\kern 10pt} t - {\left( {{\bf{\hat m}}_l^\text{R}} \right)^\text{T}}{{\bf{x}}_{\bf{R}}} \le {\left( {{\bf{\hat m}}_l^\text{F}} \right)^\text{T}}{{\bf{x}}_{\bf{F}}}, {\kern 3pt} \forall l \in {\cal L} \\
&{\kern 23pt} {x_m^{\text R}} \in {\cal X}_\text{DAC}, {\kern 3pt} \forall m = \left\{ {1,2,\cdots, N_R} \right\}
\label{eq_35}
\end{aligned}
\end{equation}
The proposed P-BB algorithm aims to update ${\bf {x}}_{\bf{R}}$ via the BB process to obtain the optimal solution of ${\cal P}_4$, while $\bf x_F$ is kept fixed throughout the algorithm.

\subsubsection{Initialization}

We select the solution obtained from the `CI 1-Bit' scheme in Section III-A as the starting point of the P-BB algorithm, and we initialize the upper bound ${\text {UB}}_0$ by substituting ${\bf{x}}_{\bf E}^{\text {CI}}$ into \eqref{eq_14}, where ${\bf{x}}_{\bf E}^{\text {CI}} = {\left[ {\Re \left( {{\bf{x}}_{\text {CI}}^\text{PSK}} \right)^\text{T},\Im \left( {{\bf{x}}_{\text {CI}}^\text{PSK}} \right)^\text{T}} \right]^\text{T}}$ represents the real representation of ${\bf x}_\text{CI}^\text{PSK}$. Accordingly, ${\text {UB}}_0$ is given by
\begin{equation}
{\text {UB}}_0 = - \mathop {\min }\limits_l \left({{\bf{m}}_l^\text{T}{\bf{x}}_{\bf{E}}^{\text {CI}}}\right).
\label{eq_33}
\end{equation}

\subsubsection{Branching}

In the branching process, we select an entry $x_n^\text{R}$ in ${\bf x}_{\bf R}$ and allocate its value. To guarantee a fast convergence speed, we adopt the adaptive subdivision rule to choose $n$ within each branching process \cite{BB}, \cite{BB2}, where $n$ satisfies:
\begin{equation}
n= \arg \mathop {\max }\limits_n \left| {x_n^\text{R} - {\cal Q}\left( {x_n^\text{R}} \right)} \right|.
\label{eq_34}
\end{equation}

Subsequently, we update $\bf x_F$ and $\bf x_R$ by removing $x_n^\text{R}$ from $\bf x_F$ and including it in $\bf x_R$, where $N_F$, $N_R$, ${\bf M_F}$ and ${\bf M_R}$ in \eqref{eq_32} are also updated accordingly. By relaxing the 1-bit constraint, the convex optimization problem to obtain the lower bound can be formulated as
\begin{equation}
\begin{aligned}
&\mathcal{P}_5: {\kern 3pt} \mathop {\min }\limits_{{{\bf {x}}_{\bf{R}}}} {\kern 3pt} - t \\
&{\kern 2pt} \text{s.t.} {\kern 10pt} t - {\left( {{\bf{\hat m}}_l^\text{R}} \right)^\text{T}}{{\bf{x}}_{\bf{R}}} \le {\left( {{\bf{\hat m}}_l^\text{F}} \right)^\text{T}}{{\bf{x}}_{\bf{F}}}, {\kern 3pt} \forall l \in {\cal L} \\
&{\kern 22pt} \left| {{x_m^{\text R}}} \right| \le \frac{1}{{\sqrt {2{N_t}} }}, {\kern 3pt} \forall m = \left\{ {1,2,\cdots, N_R} \right\}
\label{eq_35}
\end{aligned}
\end{equation}
The value of the lower bound is equal to the objective of ${\cal P}_5$ with the optimal $\bf x_R$, i.e.,
\begin{equation}
{\text {LB}}= - \mathop {\min }\limits_l \left( {{\left( {{\bf{\hat m}}_l^\text{F}} \right)^\text{T}}{{\bf{x}}_{\bf{F}}} + {\left( {{\bf{\hat m}}_l^\text{R}} \right)^\text{T}}{{\bf{x}}_{\bf{R}}}} \right),
\label{eq_36}
\end{equation}
and the corresponding upper bound is obtained by enforcing a 1-bit quantization on the resulting $\bf x_R$, given by
\begin{equation}
{\text {UB}}= - \mathop {\min }\limits_l \left( {{\left( {{\bf{\hat m}}_l^\text{F}} \right)^\text{T}}{{\bf{x}}_{\bf{F}}} + {\left( {{\bf{\hat m}}_l^\text{R}} \right)^\text{T}} {\cal Q} \left( {\bf x_R} \right) } \right).
\label{eq_37}
\end{equation}
It should be noted that $\mathcal{P}_5$ needs to be solved twice in each branching operation, since $x_n^\text{R}$ can take the value of either $-\frac{1}{\sqrt {2N_t}}$ (left child) or $\frac{1}{\sqrt {2N_t}}$ (right child). For notational convenience, we denote ${\text {LB}}^-$ (${\text {UB}}^-$) and ${\text {LB}}^+$ (${\text {UB}}^+$) as the corresponding obtained lower bound (upper bound) for the left child and right child, respectively.

\subsubsection{Bounding}

In the bounding process, we update the upper bound ${\text {UB}}_0$ and remove sub-optimal branches. To be more specific, ${\text {UB}}_0$ is updated as
\begin{equation}
{\text {UB}}_0=\min \left\{ { {\text {UB}}_0, {\kern 1pt} {\text {UB}}^- , {\kern 1pt} {\text {UB}}^+ } \right\},
\label{eq_38}
\end{equation}
and we denote ${\bf x}_{{\text {UB}}_0}$ as the 1-bit signal vector that returns ${\text {UB}}_0$. Importantly, if the value of the lower bound (${\text {LB}}^-$ or ${\text {LB}}^+$) is smaller than this updated upper bound, the corresponding obtained $\bf x_R$ is a valid branch. Otherwise, if the value of the lower bound (${\text {LB}}^-$ or ${\text {LB}}^+$) is larger than this updated upper bound, the corresponding obtained signal vector and all its subsequent branches are sub-optimal and can be excluded from the algorithm, which makes the BB process more efficient than the exhaustive search method.

\begin{figure}[!t]
\centering
\includegraphics[scale=0.4]{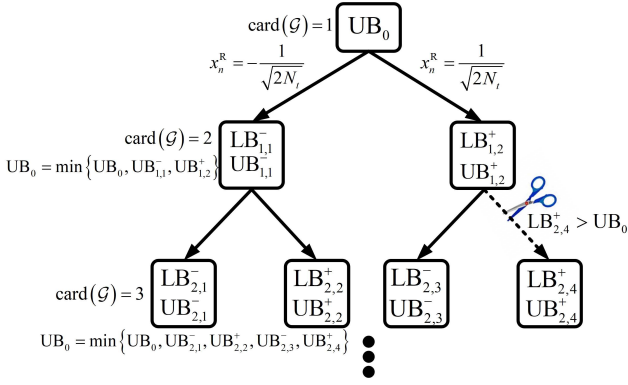}
\caption{An illustration for the BB process}
\end{figure}

\subsubsection{Algorithm}

We repeat the above branching and bounding process until all the entries in ${\bf x}_{\bf R}$ have been included in $\bf x_F$, as illustrated in Fig. 3, and the final solution of the proposed P-BB approach is obtained as the signal vector that returns the optimal upper bound value ${\text {UB}}_0$. For clarity, we summarize the above procedure in Algorithm 1 below, where ${\bf{U}} = \left[ {{{\bf{I}}_{{N_t}}}, \jmath \cdot {{\bf{I}}_{{N_t}}}} \right]$ transforms a real vector into its complex equivalence.

\begin{algorithm}
  \caption{1-Bit Precoding based on P-BB for PSK}
  \begin{algorithmic}
  
  \State {\bf Input:} ${\bf s}$, $\bf H$, ${\cal G}=\emptyset$
  \State {\bf Output:} ${\bf x}_\text{P-BB}^\text{PSK}$
  \State Obtain $\bf \tilde x_E$ by solving ${\cal P}_3$; Obtain ${\cal S}$ via \eqref{eq_26};
  \State Obtain ${\bf x}_{\bf E}^\text{CI}={\cal Q} \left( {\bf \tilde x_E} \right)$;
  \State Denote ${\bf x}^{(1)}={\bf x}_{\bf E}^\text{CI}$; Set ${\cal G}=\left\{ {{\bf x}^{(1)}} \right\}$; Obtain ${\text {UB}}_0$ via \eqref{eq_33};
  \For {$i = 1:\text{card}\left( {\cal S} \right)$}
  \For {$m = 1:\text{card}\left( {\cal G} \right)$}
    \State ${{\bf{x}}_\text{temp}} \gets {{\bf{x}}^{(m)}}$;
    \State Decompose ${{\bf{x}}_\text{temp}}$ into $\bf x_F$ and $\bf x_R$ via \eqref{eq_31};
    \State Find $n$ via \eqref{eq_34};
    \State {\bf Branching}
    \State {\bf Left Child:}
    \State $x_n^\text{R} \gets -\frac{1}{\sqrt{2N_t}}$; Obtain ${\bf x}_{\bf R}^-$ by solving ${\cal P}_5$;
    \State Obtain ${\text {LB}}^-$ and ${\text {UB}}^-$ via \eqref{eq_36} and \eqref{eq_37};
    \State {\bf Right Child:}
    \State $x_n^\text{R} \gets \frac{1}{\sqrt{2N_t}}$; Obtain ${\bf x}_{\bf R}^+$ by solving ${\cal P}_5$;
    \State Obtain ${\text {LB}}^+$ and ${\text {UB}}^+$ via \eqref{eq_36} and \eqref{eq_37};
      
    \State {\bf Bounding}
    \State Update ${\text {UB}}_0$ via \eqref{eq_38}; Update ${\bf x}_{{\text {UB}}_0}$;
    \State Set ${\cal G}=\emptyset$;
    \If {$\text{LB}^- < {\text {UB}}_0$}
    \State Include ${\bf x}^{\left({\text{card}\left({\cal G}\right)+1}\right)} = {\left[ {{\bf{x}}_{\bf{F}}^\text{T},{{\left( {{\bf{x}}_{\bf{R}}^- } \right)}^\text{T}}} \right]^\text{T}}$ in $\cal G$;
    \EndIf
    \If {$\text{LB}^+ < {\text {UB}}_0$}
    \State Include ${\bf x}^{\left({\text{card}\left({\cal G}\right)+1}\right)} = {\left[ {{\bf{x}}_{\bf{F}}^\text{T},{{\left( {{\bf{x}}_{\bf{R}}^+ } \right)}^\text{T}}} \right]^\text{T}}$ in $\cal G$;
    \EndIf
  \EndFor
  \EndFor
\State Output ${\bf x}_\text{P-BB}^\text{PSK} = {\bf U} {\bf x}_{{\text {UB}}_0}$.
   \end{algorithmic}
\end{algorithm}

\subsection{A Low-Complexity Alternative via OPSU}
While the proposed P-BB algorithm exhibits a significant complexity reduction compared to the F-BB method, it may still need to search the entire subspace ${\cal X}_{\text {DAC}}^{N_R}$ in the worst case, which may not be favorable when the number of users is large. Therefore, in this section we further introduce a low-complexity alternative approach based on an `ordered partial sequential update' (OPSU) process, which is essentially a greedy algorithm. Firstly, it is observed in Algorithm 1 that, when updating $\bf x_F$, the P-BB approach considers the effects of all the residual entries in ${\bf x}_{\bf R}$ on the resulting $\bf \Lambda$ by solving ${\cal P}_5$. To pursue a more computationally-efficient approach, we propose a sub-optimal procedure by only considering the effect of a single entry in ${\bf x}_{\bf R}$ at a time on the objective function. Because of this design, the sequence how we select ${\bf x}_{\bf R}$ each time may further have an effect on the solution of $\bf x_E$ and lead to different local optimums. 

To be more specific, we first rewrite $\bf \Lambda$ as
\begin{equation}
{\bf \Lambda}={{\bf{M}}_{\bf{F}}}{{\bf{x}}_{\bf{F}}} + \sum\limits_{k = 1}^{N_R} {{\bf{m}}_k^{\bf{R}} x_k^{\text R}},
\label{eq_39}
\end{equation}
where ${\bf{m}}_k^{\bf{R}}$ represents the $k$-th column in $\bf M_R$, and ${{\bf{M}}_{\bf{R}}} = \left[ {{\bf{m}}_1^{\bf{R}},{\bf{m}}_2^{\bf{R}}, \cdots ,{\bf{m}}_{N_R}^{\bf{R}}} \right]$. For the proposed `OPSU' approach, in each iteration we aim to choose the value for ${x}_k^{\text R}$ that can increase the value of the minimum entry in $\bf \Lambda$, while keeping other entries in ${\bf \hat x}_{\bf E}$ fixed. Meanwhile, we note that the amplitudes of the entries in the corresponding ${\bf m}_k^{\bf R}$ also have an effect on the resulting ${\bf \Lambda}$. Therefore, by denoting
\begin{equation}
{\gamma _k} = \mathop {\min }\limits_l \left( {\left| {{\bf{m}}_k^{\bf{R}}\left( l \right)} \right|} \right),
\label{eq_40}
\end{equation}
we propose to first allocate values for ${x}_k^{\text R}$ whose corresponding ${\bf m}_k^{\bf R}$ has the most significant impact on $\bf \Lambda$, i.e., ${\bf m}_k^{\bf R}$ that has the largest value of $\gamma_k$. We repeat the above process until all the entries in ${\bf x}_{\bf R}$ have been visited, and this iterative process is summarized in Algorithm 2, where ${\text {sort}}\left[  \cdot  \right]$ is the sort function following a descending order.

\begin{algorithm}
  \caption{1-Bit Precoding based on OPSU for PSK}
  \begin{algorithmic}
  
  \State {\bf Input:} ${\bf s}$, $\bf H$
  \State {\bf Output:} ${\bf x}_\text{OPSU}^\text{PSK}$
  \State Obtain $\bf M$ based on ${\bf s}$ and $\bf H$; Calculate each $\gamma_k$ via \eqref{eq_38};
  \State Obtain $\bf \tilde x_E$ by solving ${\cal P}_3$; Obtain ${\cal S}$ via \eqref{eq_26};  
  \State Decompose $\bf \tilde x_E$ into ${\bf x_F}$ and ${\bf x}_{\bf R}$ via \eqref{eq_31};  
  \State Obtain ${\text {UB}}_0$ via \eqref{eq_33} based on ${\bf x}_{\bf E}^\text{CI}={\cal Q}\left( {\bf \tilde x_E} \right)$;
  \State Construct ${\bm \gamma}=\left[ {\gamma_1,\gamma_2,\cdots,\gamma_{\text{card}\left({\cal S}\right)}} \right]^\text{T}$;
  \State Obtain ${\bm \gamma}_{\text O}={\text {sort}}\left[  {\bm \gamma}  \right]=\left[ {\gamma_1^{\text O},\gamma_2^{\text O},\cdots,\gamma_{\text{card}\left({\cal S}\right)}^{\text O}} \right]^\text{T}$;
  \For {$i=1:\text{card}\left( {\cal S} \right)$}
    \State Find the index $k$ in \eqref{eq_38} where $\gamma_k = \gamma_i^{\text O}$;
    \State {\bf Left Child:}
      \State $ x_k^{\text R} \gets -\frac{1}{\sqrt{2N_t}}$; Calculate ${\text{UB}}^-$;
    \State {\bf Right Child:}
      \State $ x_k^{\text R} \gets \frac{1}{\sqrt{2N_t}}$; Calculate ${\text{UB}}^+$;
      \State Update ${\text {UB}}_0$ via \eqref{eq_38}; Update ${\bf x}_{{\text {UB}}_0}$;
  
  \EndFor
  \State Output ${\bf x}_\text{OPSU}^\text{PSK}={\bf U} {\bf x}_{{\text {UB}}_0}$.
  
   \end{algorithmic}
\end{algorithm}

Compared to the P-BB method proposed in the previous section where the cardinality of the set $\cal G$ in Algorithm 1 may keep increasing after each iteration, the major complexity gain for the low-complexity `OPSU' method proposed in this section comes from the fact that we only consider one feasible solution ${\bf \tilde x_E}$ and update its entries following an iterative manner. In this case, $\text{card}\left( {\cal G} \right)=1$ and therefore the inner iterative process in Algorithm is no longer required. Another complexity reduction comes from the fact that the proposed `OPSU' method avoids the need to solve the optimization problem ${\cal P}_5$ within each iteration. Both of the above make the `OPSU' method more computationally efficient than the P-BB approach.

\section{1-Bit Precoding for QAM Signaling}
In this section, we focus on CI-based 1-bit precoding approaches when QAM signaling is considered at the BS. In this case, the received signal vector $\bf y$ needs to be further re-scaled for correct demodulation, expressed as
\begin{equation}
{\bf{r}} = \beta  \cdot {\bf{y}} = \beta  \cdot {\bf{Hx}} + \beta  \cdot {\bf{n}},
\label{eq_41}
\end{equation}
where $\bf r$ is the received symbol vector for demodulation, and $\beta$ is the precoding factor that can be obtained by minimizing the MSE between $\bf r$ and $\bf s$, given by \cite{dac11}
\begin{equation}
\beta  = \frac{{\Re \left( {{{\bf{x}}^\text{H}}{{\bf{H}}^\text{H}}{\bf{s}}} \right)}}{{\left\| {{\bf{Hx}}} \right\|_2^2 + K{\sigma ^2}}}.
\label{eq_42}
\end{equation}

\subsection{CI Condition and Problem Formulation}

\begin{figure}[!b]
\centering
\includegraphics[scale=0.6]{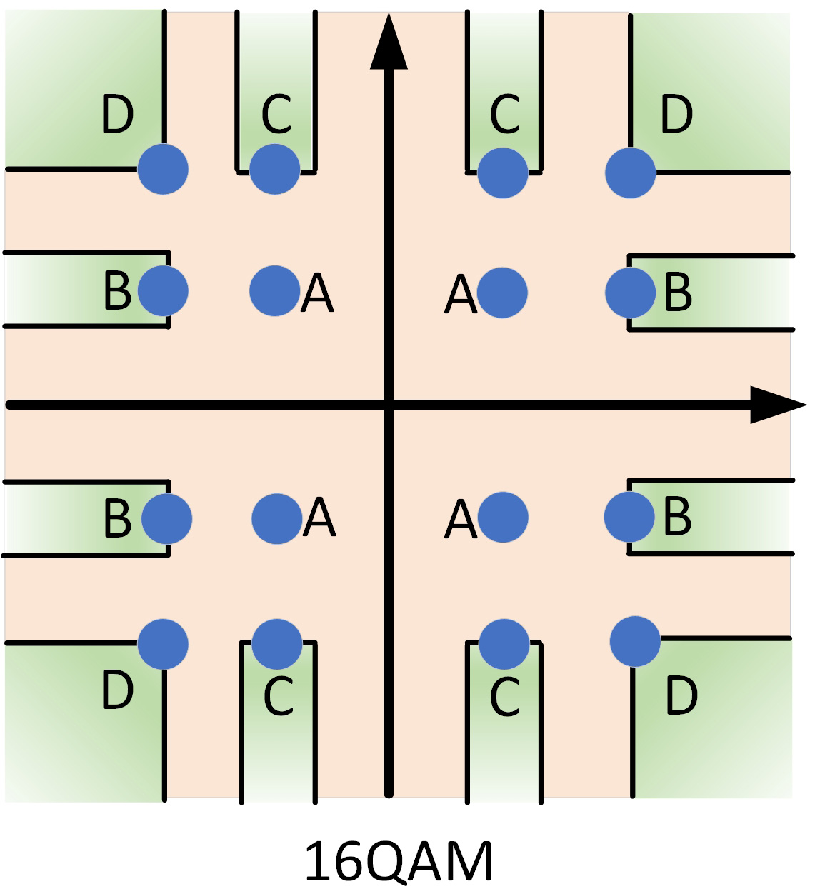}
\caption{An illustrative example of CI condition for QAM}
\end{figure}

Similar to Section III, we begin by considering the CI-based 1-bit precoding design for QAM, where we still decompose the symbol and noiseless received signal following \eqref{eq_5} and \eqref{eq_6}. In the case where QAM constellations are considered, the expressions for $s_k^{\cal A}$ and $s_k^{\cal B}$ can be simplified into
\begin{equation}
s_k^{\cal A}= \Re \left( {s_k} \right), {\kern 3pt} s_k^{\cal B}=\jmath \cdot \Im \left( {s_k} \right), {\kern 3pt} \forall k \in {\cal K}.
\label{eq_43}
\end{equation}
For the mathematical CI condition for QAM constellations, we follow \cite{ci13} and consider the multi-user interference on the inner constellation points as only destructive, as illustrated in Fig. 4, where a 16QAM constellation is depicted. Accordingly, we divide the real scalars $\alpha_k^{\cal A}$ and $\alpha_k^{\cal B}$, $\forall k \in {\cal K}$ into two groups $\cal O$ and $\cal I$, where the entries in $\cal O$ correspond to the real or imaginary part of the symbols that can exploit CI, i.e., both the real and imaginary part of the constellation point `D', the real part of `B' and the imaginary part of `C', as shown in Fig. 4, while $\cal I$ consists of the residual entries corresponding to the symbols that cannot benefit from CI. We can then obtain
\begin{equation}
{\cal O} \cup {\cal I} = \left\{ {\alpha _1^{\cal A},\alpha _1^{\cal B},\alpha _2^{\cal A},\alpha _2^{\cal B}, \cdots ,\alpha _K^{\cal A},\alpha _K^{\cal B}} \right\},
\label{eq_44}
\end{equation}
and
\begin{equation}
\text{card}\left( {\cal O} \right) + \text{card}\left( {\cal I} \right) = 2K.
\label{eq_45}
\end{equation}
Subsequently, the CI condition for QAM constellation points can be expressed as
\begin{equation}
\alpha_{m}^{\cal O} \ge \alpha_{n_1}^{\cal I}, {\kern 3pt} \alpha_{n_1}^{\cal I}=\alpha_{n_2}^{\cal I}, {\kern 3pt} \forall \alpha_{m}^{\cal O} \in {\cal O},{\kern 3pt} \forall \alpha_{n_1}^{\cal I}, \alpha_{n_2}^{\cal I} \in {\cal I},
\label{eq_46}
\end{equation}
and the corresponding 1-bit precoding problem that exploits CI can be formulated as
\begin{equation}
\begin{aligned}
&\mathcal{P}_6: {\kern 3pt} \mathop {\max }\limits_{\bf{x}} {\kern 3pt} t \\
&{\kern 2pt} \text{s.t.} {\kern 10pt} {{\bf{h}}_k^\text{T}}{\bf x} = \alpha_k^{\cal A} s_k^{\cal A} + \alpha_k^{\cal B} s_k^{\cal B}, {\kern 3pt} \forall k \in {\cal K} \\
&{\kern 24pt} \alpha _m^{\cal O} \ge t, {\kern 3pt} \forall \alpha _m^{\cal O} \in {\cal O} \\
&{\kern 24pt} \alpha _n^{\cal I} = t, {\kern 3pt} \forall \alpha _m^{\cal O} \in {\cal I} \\
&{\kern 24pt} x_n \in {\cal X}_{{\text{DAC}}}, {\kern 3pt} \forall n \in {\cal N}
\label{eq_47}
\end{aligned}
\end{equation}
{\bf Remark:} ${\cal P}_6$ is a non-convex optimization problem. More importantly, due to the fact that the equality constraints and the 1-bit constraints cannot both be satisfied at the same time in general, we note that the original optimization problem ${\cal P}_6$ for QAM constellations is an infeasible problem in nature, as opposed to ${\cal P}_1$ formulated for PSK which is always feasible. This infeasibility also makes the P-BB algorithm designed for PSK signaling not directly applicable.

To obtain a feasible 1-bit solution for QAM signaling, we can relax the 1-bit constraint in ${\cal P}_6$ by following similar steps as in \eqref{eq_9} and \eqref{eq_10}, where the relaxed optimization problem can be formulated as
\begin{equation}
\begin{aligned}
&\mathcal{P}_7: {\kern 3pt} \mathop {\max }\limits_{\bf \tilde{x}} {\kern 3pt} t \\
&{\kern 2pt} \text{s.t.} {\kern 10pt} {{\bf{h}}_k^\text{T}}{\bf \tilde x} = \alpha_k^{\cal A} s_k^{\cal A} + \alpha_k^{\cal B} s_k^{\cal B}, {\kern 3pt} \forall k \in {\cal K} \\
&{\kern 24pt} \alpha _m^{\cal O} \ge t, {\kern 3pt} \forall \alpha _m^{\cal O} \in {\cal O} \\
&{\kern 24pt} \alpha _n^{\cal I} = t, {\kern 3pt} \forall \alpha _m^{\cal O} \in {\cal I} \\
&{\kern 22pt} \left| {\Re \left( {{{\tilde x}_n}} \right)} \right| \le \frac{1}{{\sqrt {2{N_t}} }}, {\kern 3pt} \forall n \in {\cal N}\\
&{\kern 22pt} \left| {\Im \left( {{{\tilde x}_n}} \right)} \right| \le \frac{1}{{\sqrt {2{N_t}} }}, {\kern 3pt} \forall n \in {\cal N}
\label{eq_48}
\end{aligned}
\end{equation}
The error rate performance for `CI 1-Bit' following this relaxation-normalization procedure serves as an upper bound of the proposed 1-bit precoding methods based on P-BB and OPSU introduced in the following. For notational convenience, we denote the obtained quantized signal vector based on this conventional CI approach as ${\bf x}_\text{CI}^\text{QAM}$.

\subsection{Analytical Study of 1-Bit CI Precoding for QAM}

In this section, we show that the results revealed in {\bf Proposition 1} for PSK signaling directly extend to QAM signaling, while the problem formulations are different. To begin with, we expand \eqref{eq_41} into its real representation, given by
\begin{equation}
{{\bf{r}}_{\bf{E}}} = {\beta} \cdot {{\bf{y}}_{\bf{E}}} = {\beta } \cdot {{\bf{H}}_{\bf{E}}}{{\bf{x}}_{\bf{E}}} + {\beta} \cdot {\bf n_E},
\label{eq_49}
\end{equation}
where ${{\bf{r}}_{\bf{E}}}$, ${{\bf{y}}_{\bf{E}}}$ and ${{\bf{n}}_{\bf{E}}}$ are the real representations of $\bf r$, $\bf y$ and $
\bf n$, respectively, similar to $\bf x_E$ as shown in \eqref{eq_12}. $\beta$ in \eqref{eq_42} can then be equivalently expressed as
\begin{equation}
\beta = \frac{{{\bf{x}}_{\bf{E}}^\text{T}{\bf{H}}_{\bf{E}}^\text{T}{{\bf{s}}_{\bf{E}}}}}{{\left\| {{{\bf{H}}_{\bf{E}}}{{\bf{x}}_{\bf{E}}}} \right\|_2^2 + K{\sigma ^2}}},
\label{eq_50}
\end{equation}
where ${\bf s_E}=\left[ {\Re \left( {{{\bf{s}}^\text{T}}} \right),\Im \left( {{{\bf{s}}^\text{T}}} \right)} \right]^\text{T}$. Following the steps in Section III-A, the relaxed optimization problem $\mathcal{P}_7$ can be equivalently transformed into 
\begin{equation}
\begin{aligned}
&\mathcal{P}_8: {\kern 3pt} \mathop {\max }\limits_{{\bf \tilde x}_{\bf E}} {\kern 3pt} t \\
&{\kern 2pt} \text{s.t.} {\kern 10pt} t \le {\left( {{\bf{m}}_m^{\cal O}} \right)^\text{T}}{{\bf{\tilde x}}_{\bf{E}}}, {\kern 3pt} \forall \alpha_m^{\cal O} \in {\cal O} \\
&{\kern 24pt} t = {\left( {{\bf{m}}_n^{\cal I}} \right)^\text{T}}{{\bf{\tilde x}}_{\bf{E}}}, {\kern 3pt} \forall \alpha_n^{\cal I} \in {\cal I} \\
&{\kern 22pt} \left| {{{\tilde x}_m^{\text E}}} \right| \le \frac{1}{{\sqrt {2{N_t}} }}, {\kern 3pt} \forall m \in {\cal M}
\label{eq_51}
\end{aligned}
\end{equation}
Based on the formulation of ${\cal P}_8$, the following proposition is presented.

{\bf Proposition 2:} Similar to the case of PSK, for ${\bf \tilde x}_{\bf E}$ obtained by solving ${\cal P}_8$, there are at least $\left( {2N_t-2K+1} \right)$ entries that already satisfy the 1-bit constraint.

{\bf Proof:} By constructing the KKT conditions of ${\cal P}_8$, this proposition can be similarly proven by contradiction following \eqref{eq_23}-\eqref{eq_30}, which is omitted here for brevity. ${\kern 60pt} \blacksquare$

\subsection{1-Bit Precoding Design via Partial Branch-and-Bound}

In this section, we propose the 1-bit precoding scheme via P-BB for QAM modulations. Before we proceed, we note that due to the infeasibility of ${\cal P}_6$ in nature as discussed in {\bf Remark}, the P-BB algorithm designed for PSK constellations cannot be directly extended to the case of QAM modulations, since the sub-problem included in the BB process will be infeasible when the number of entries in $\bf x_R$ that are to be optimized is smaller than $\text{card}\left( {\cal I} \right)$. To circumvent this issue, when we design the P-BB algorithm for QAM signaling after obtaining ${\bf \tilde x}_{\bf E}$, we consider the MSE criterion as the objective function instead, which is defined as
\begin{equation}
{\text {MSE}}=\left\| {{{\bf{s}}_{\bf{E}}} - {\beta} \cdot {{\bf{H}}_{\bf{E}}}{\bf x_E}} \right\|_2^2 + \beta^2K{\sigma ^2}.
\label{eq_52}
\end{equation}

Based on the expression for MSE as shown in \eqref{eq_52}, we note another distinct feature when QAM signaling is considered: Compared to PSK signaling in which case the objective function only includes $\bf x_E$, the objective function for QAM signaling also includes the precoding factor $\beta$, which is a function of the transmit signal vector $\bf x_E$ that is to be optimized. The non-linear relationship between $\beta$ and $\bf x_E$, as observed in \eqref{eq_50}, makes the direct minimization on MSE difficult to solve. Nevertheless, noting that $\beta$ and $\bf x_E$ are uncoupled in the expression for MSE, the alternating optimization framework can be adopted as an effective method \cite{Alt}. To be more specific, the alternating optimization selects ${\bf x}_\text{CI}^\text{QAM}$ as the starting point, and iteratively update $\beta$ and $\bf x_E$ until convergence, where the update for $\beta$ follows \eqref{eq_50}, and the updated $\bf x_E$ is obtained by minimizing the MSE in \eqref{eq_52}, given by
\begin{equation}
\begin{aligned}
&\mathcal{P}_{9}: {\kern 3pt} \mathop {\min }\limits_{{{\bf{x}}_{\bf{E}}}} \left\| {{{\bf{s}}_{\bf{E}}} - {\beta} \cdot {{\bf{H}}_{\bf{E}}}{\bf x_E}} \right\|_2^2\\
&{\kern 2pt} \text{s.t.} {\kern 10pt} x_m^\text{E} \in {\cal X}_{\text {DAC}}, {\kern 3pt} \forall m \in {\cal M}
\label{eq_53}
\end{aligned}
\end{equation}
where we note that the term $\beta^2K{\sigma ^2}$ is constant when $\beta$ is fixed, which is therefore omitted.

For clarity, we first summarize the main steps of the alternating optimization framework in Algorithm 3 before proceeding, where $\epsilon_0$ is a pre-defined threshold for convergence.

\begin{algorithm}
  \caption{Alternating Optimization Framework for 1-Bit Precoding based on P-BB for QAM}
  \begin{algorithmic}
  
  \State {\bf Input:} ${\bf s}$, $\bf H$, $\sigma^2$, $\epsilon_0$
  \State {\bf Output:} ${\bf x}_{\text {P-BB}}^\text{QAM}$
  \State Obtain $\bf M$ based on ${\bf s}$ and $\bf H$;
  \State Obtain $\bf \tilde x_E$ by solving ${\cal P}_8$; Express ${\bf x}_{\bf E}={\cal Q}\left({\bf \tilde x_E}\right)$;
  \State Calculate $\beta$ based on ${\bf x}_{\bf E}$ via \eqref{eq_50};
  \State Calculate $\text{MSE}_0$ based on $\beta$ and ${\bf x}_{\bf E}$ via \eqref{eq_52};  
  \While {$\epsilon>\epsilon_0$}
    \State Update $\beta$ via \eqref{eq_50} based on $\bf x_E$;
    \State Update $\bf x_E$ by solving ${\cal P}_{9}$ via P-BB with the given $\beta$;
    \State Calculate MSE based on the given $\beta$ and the updated $\bf x_E$ via \eqref{eq_52};
    \State $\epsilon = \left| {\text{MSE} - \text{MSE}_0} \right|$;
    \State $\text{MSE}_0 \gets \text{MSE}$;
  \EndWhile
  \State Output ${\bf x}_{\text {P-BB}}^\text{QAM}={\bf U} {\bf x_E}$.
   \end{algorithmic}
\end{algorithm}

In the following, we briefly describe the P-BB process within the alternating optimization framework for QAM signaling, which generally follows the P-BB process for PSK in Section III-A. The major difference lies in the formulated optimization problems for obtaining $\bf x_R$ and the corresponding calculation of the lower bounds and upper bounds, since the criterion has switched to MSE minimization.

\subsubsection{Initialization}

The initial upper bound $\text{UB}_0$ can be obtained based on the expression for MSE in \eqref{eq_52} as
\begin{equation}
\text{UB}_0=\left\| {{{\bf{s}}_{\bf{E}}} - {\beta} \cdot {{\bf{H}}_{\bf{E}}}{\bf x_E}} \right\|_2^2 + \beta^2K{\sigma ^2},
\label{eq_54}
\end{equation}
where we note that $\beta$ is fixed in this BB process due to the alternating optimization approach.

\subsubsection{Branching}
Similar to the case for PSK, we rearrange $\bf \tilde x_E$ obtained from solving ${\cal P}_8$ into $\bf \hat x_E$ such that $\bf \hat x_E$ can be decomposed as in \eqref{eq_31}, where $\bf x_F$ and $\bf x_R$ are similarly defined. To proceed, we select an entry in $\bf x_R$ and allocate its value following the adaptive subdivision rule. The resulting optimization problem on $\bf x_R$ that minimizes the MSE can then be constructed as
\begin{equation}
\begin{aligned}
&\mathcal{P}_{10}: {\kern 3pt} \mathop {\min }\limits_{\bf x_R} \left\| {{{\bf{s}}_{\bf{E}}} - {\beta} \cdot {{\bf \hat H}_{\bf{E}}} {\left[ {{\bf{x}}_{\bf{F}}^\text{T},{\bf{x}}_{\bf{R}}^\text{T}} \right]^\text{T}} } \right\|_2^2\\
&{\kern 4pt} \text{s.t.} {\kern 12pt} x_m^\text{R} \in {\cal X}_{\text {DAC}}, {\kern 3pt} \forall m \in {\cal M}
\label{eq_55}
\end{aligned}
\end{equation}
where $\bf \hat H_E$ denotes $\bf H_E$ with the corresponding column rearrangement. By decomposing $\bf \hat H_E$ into ${{\bf{\hat H}}_{\bf{E}}} = \left[ {{{\bf{H}}_{\bf{F}}},{{\bf{H}}_{\bf{R}}}} \right]$, the objective function of ${\cal P}_{10}$ can be simplified into
\begin{equation}
\begin{aligned}
&\left\| {{{\bf{s}}_{\bf{E}}} - {\beta} \cdot \left[ {{{\bf{H}}_{\bf{F}}},{{\bf{H}}_{\bf{R}}}} \right] {\left[ {{\bf{x}}_{\bf{F}}^\text{T},{\bf{x}}_{\bf{R}}^\text{T}} \right]^\text{T}} } \right\|_2^2 \\
= &\left\| {\left( {{{\bf{s}}_{\bf{E}}} - \beta  \cdot {{\bf{H}}_{\bf{F}}}{{\bf{x}}_{\bf{F}}}} \right) - \beta  \cdot {{\bf{H}}_{\bf{R}}}{{\bf{x}}_{\bf{R}}}} \right\|_2^2 \\
= &\left\| {{{\bf{s}}_{\bf{F}}} - \beta  \cdot {{\bf{H}}_{\bf{R}}}{{\bf{x}}_{\bf{R}}}} \right\|_2^2,
\end{aligned}
\label{eq_56}
\end{equation} 
where we introduce ${\bf s_F}= {{{\bf{s}}_{\bf{E}}} - \beta  \cdot {{\bf{H}}_{\bf{F}}}{{\bf{x}}_{\bf{F}}}}$ that is fixed within the P-BB process for a given $\beta$. To obtain the lower bound, we relax the 1-bit constraint in ${\cal P}_{10}$ to arrive at a convex least-squares (LS) problem as
\begin{equation}
\begin{aligned}
&\mathcal{P}_{11}: {\kern 3pt} \mathop {\min }\limits_{\bf x_R} \left\| {{{\bf{s}}_{\bf{F}}} - \beta  \cdot {{\bf{H}}_{\bf{R}}}{{\bf{x}}_{\bf{R}}}} \right\|_2^2\\
&{\kern 4pt} \text{s.t.} {\kern 10pt} \left| {{x_m^{\text R}}} \right| \le \frac{1}{{\sqrt {2{N_t}} }}, {\kern 3pt} \forall m = \left\{ {1,2,\cdots, N_R} \right\}
\label{eq_55}
\end{aligned}
\end{equation}
The lower bound is equal to the objective of ${\cal P}_{11}$ with the optimal $\bf x_R$, i.e.,
\begin{equation}
\text{LB}=\left\| {{{\bf{s}}_{\bf{F}}} - \beta  \cdot {{\bf{H}}_{\bf{R}}}{{\bf{x}}_{\bf{R}}}} \right\|_2^2 + \beta^2K{\sigma ^2},
\label{eq_56}
\end{equation}
and the corresponding upper bound is obtained by enforcing the 1-bit quantization on the optimal $\bf x_R$, given by
\begin{equation}
\text{UB}=\left\| {{{\bf{s}}_{\bf{F}}} - \beta  \cdot {{\bf{H}}_{\bf{R}}} {\cal Q}\left({{\bf{x}}_{\bf{R}}}\right)} \right\|_2^2 + \beta^2K{\sigma ^2}.
\label{eq_57}
\end{equation}
Similar to the case of PSK, ${\cal P}_{11}$ needs to be solved twice, for the left child and right child, respectively.

\subsubsection{Bounding and Algorithm}
The bounding operation and the algorithm for QAM signaling generally follow the bounding process for PSK signaling and Algorithm 1, which are therefore omitted for brevity.

By substituting the P-BB algorithm into the alternating optimization framework in Algorithm 3, the final 1-bit solution based on P-BB for QAM signaling can then be obtained.

\subsection{A Low-Complexity Alternative}
In this section, we further develop a low-complexity alternative method based on OPSU for QAM signaling as well. Similar to the case of PSK, we consider the sub-optimal approach where we update a single entry in $\bf x_R$ at a time following an iterative process. Following the P-BB method proposed for QAM signaling, we adopt the MSE metric when designing this low-complexity algorithm. 

To be more specific, we firstly calculate the initial precoding factor $\beta_{\text 0}$ and ${\text {MSE}}_0$ based on the obtained $\bf \tilde x_E$ by solving ${\cal P}_8$. Subsequently, in each iteration we allocate the value ($-\frac{1}{\sqrt{2N_t}}$ or $\frac{1}{\sqrt{2N_t}}$) for one entry in ${\bf x}_{\bf R}$, calculate the corresponding precoding factor $\beta$, and choose the one that returns a lower MSE value, bearing in mind that the entry in ${\bf x}_{\bf R}$ with a larger value of the corresponding $\gamma_k$ in \eqref{eq_40} has a higher priority to be considered, as in the case for PSK signaling. This process is repeated until all the entries in ${\bf x}_{\bf R}$ have been visited, and the above iterative approach is summarized in Algorithm 4 below.

\begin{algorithm}
  \caption{1-Bit Precoding based on OPSU for QAM}
  \begin{algorithmic}
  
  \State {\bf Input:} ${\bf s}$, $\bf H$
  \State {\bf Output:} ${\bf x}_\text{OPSU}^\text{QAM}$
  \State Obtain $\bf M$ based on ${\bf s}$ and $\bf H$; Calculate each $\gamma_k$ via \eqref{eq_38};
  \State Obtain $\bf \tilde x_E$ by solving ${\cal P}_8$; Express ${\bf x}_{\bf E}^\text{CI}={\cal Q}\left( {\bf \tilde x_E} \right)$; 
  \State Obtain ${\cal S}$ via \eqref{eq_26}; Decompose $\bf \tilde x_E$ via \eqref{eq_31}; 
  \State Obtain $\beta_0$ via \eqref{eq_50} and ${\text {MSE}}_0$ via \eqref{eq_52} based on ${\bf x}_{\bf E}^\text{CI}$;
  \State Construct ${\bm \gamma}=\left[ {\gamma_1,\gamma_2,\cdots,\gamma_{\text{card}\left({\cal S}\right)}} \right]^\text{T}$;
  \State Obtain ${\bm \gamma}_{\text O}={\text {sort}}\left[  {\bm \gamma}  \right]=\left[ {\gamma_1^{\text O},\gamma_2^{\text O},\cdots,\gamma_{\text{card}\left({\cal S}\right)}^{\text O}} \right]^\text{T}$;
  \For {$i=1:\text{card}\left( {\cal S} \right)$}
    \State Find the index $k$ in \eqref{eq_38} where $\gamma_k = \gamma_i^{\text O}$;
    \State {\bf Left Child:}
      \State $ x_k^{\text R} \gets -\frac{1}{\sqrt{2N_t}}$; Calculate $\beta^-$ and ${\text{MSE}}^-$;
    \State {\bf Right Child:}
      \State $ x_k^{\text R} \gets \frac{1}{\sqrt{2N_t}}$; Calculate $\beta^+$ and ${\text{MSE}}^+$;
      \State Update ${\text {MSE}}_0=\min \left\{ { \text{MSE}_0, {\kern 1pt} {\text {MSE}}^- , {\kern 1pt} {\text {MSE}}^+ } \right\}$; 
      \State Update ${\bf x}_{{\text {UB}}_0}$;
  
  \EndFor
  \State Output ${\bf x}_\text{OPSU}^\text{QAM}={\bf U} {\bf x}_{{\text {UB}}_0}$.
  
   \end{algorithmic}
\end{algorithm}

\section{Numerical Results}
In this section, numerical results of the proposed approaches are presented based on Monte Carlo simulations. In each plot, the transmit SNR is defined as $\rho  = \frac{1}{\sigma^2}$, where we have assumed unit transmit power. We compare our proposed methods with quantized linear and non-linear precoding approaches in the literature. For clarity, the following abbreviations are used throughout this section:

\begin{enumerate}
\item `ZF Inf-Bit': Unquantized ZF precoding with infinite-precision DACs;
\item `ZF 1-Bit': 1-bit quantized ZF approach;
\item `MMSE 1-Bit': 1-bit quantized MMSE-based approach \cite{dac4};
\item `GDM': The 1-bit gradient descend method \cite{dac7};
\item `DP': The direct perturbation method for QPSK with iteration number $\text{N}_\text{DP}$ \cite{dac16};
\item `C1PO': The C1PO algorithm with iteration number $\text{N}_\text{C1PO}$ \cite{dac8};
\item `C2PO': The C2PO algorithm with iteration number $\text{N}_\text{C2PO}$ \cite{dac9};
\item `CI 1-Bit': The 1-bit CI-based method by quantizing the solution of ${\cal P}_2$ for PSK or ${\cal P}_8$ for QAM;
\item `CI 1-Bit OPSU': The proposed 1-bit OPSU method;
\item `CI 1-Bit P-BB': The proposed 1-bit P-BB method;
\item `1-Bit F-BB': The optimal F-BB method \cite{dac14}.
\end{enumerate}

\subsection{Results for PSK}

\begin{figure}[!t]
\centering
\subfloat[$N_t=8$, $K=2$]
{%
  \includegraphics[clip,width=0.9\columnwidth]{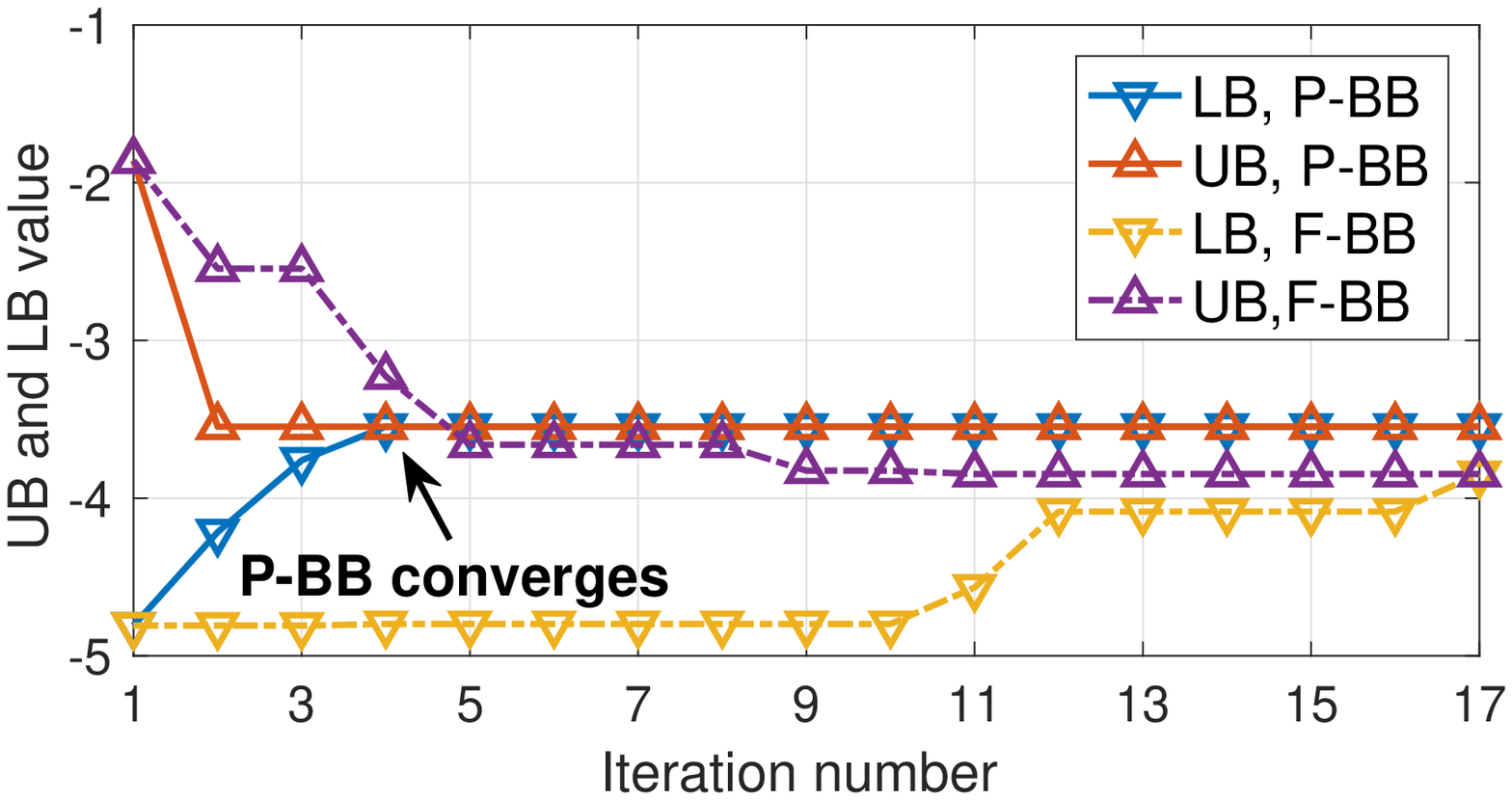}%
}

\subfloat[$N_t=64$, $K=16$]
{%
  \includegraphics[clip,width=0.9\columnwidth]{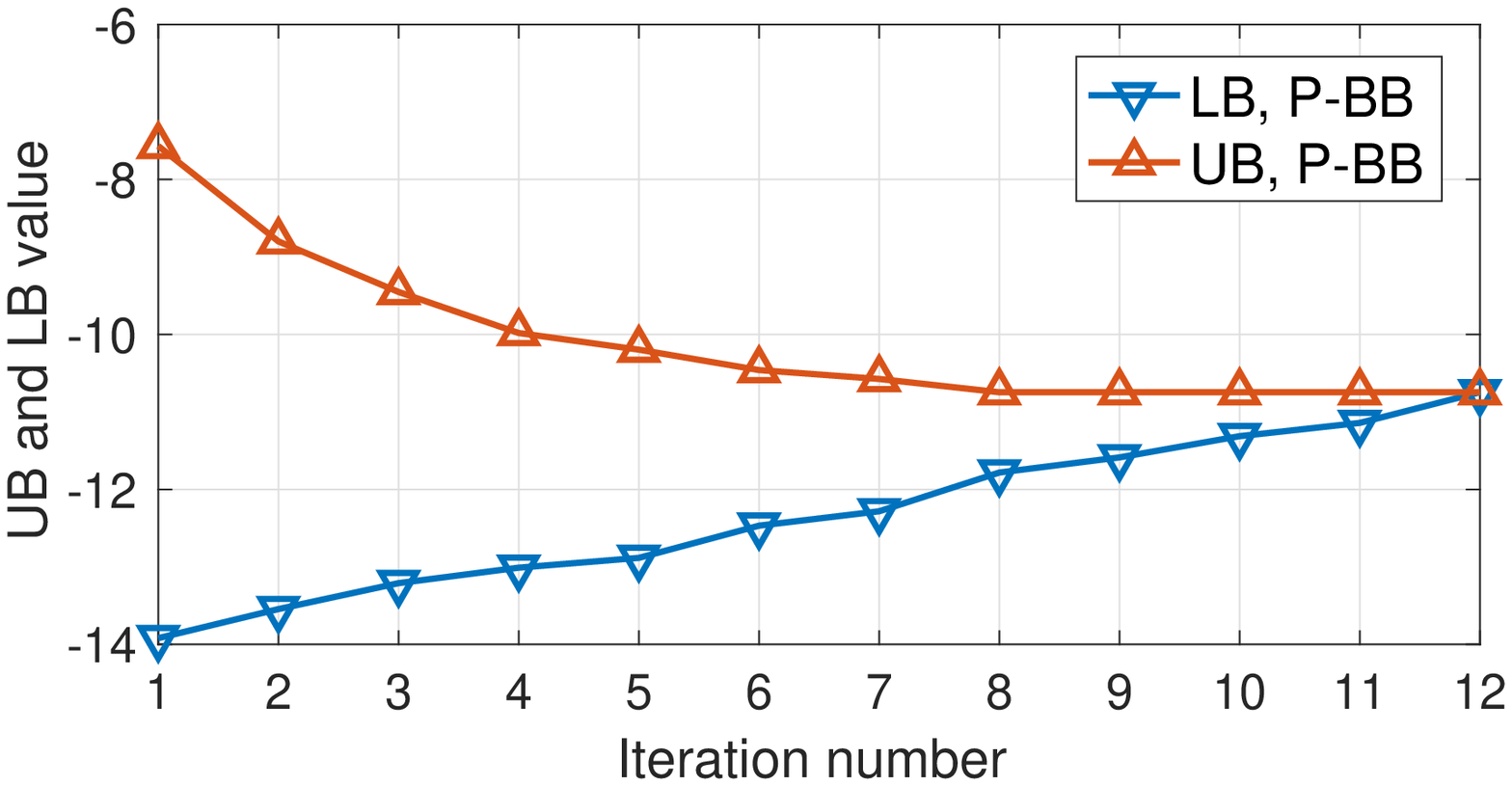}%
}
\caption{Convergence of the P-BB method, QPSK}
\end{figure}

Before presenting the BER results, we firstly show the convergence of the proposed P-BB method, as depicted in Fig. 5, where QPSK modulation is adopted. In a small-scale $8 \times 2$ MIMO system as in Fig. 5 (a), the convergence of both the F-BB method in \cite{dac14} and the P-BB method proposed in this paper is presented. Compared to the F-BB method that requires 17 iterations to converge, we observe that the proposed P-BB method converges within only 4 iterations. We can also observe that the performance gap between the optimal F-BB approach and the proposed P-BB scheme is marginal, as will also be shown by the BER result in the following. When the MIMO system scales up to $64 \times 16$, as depicted in Fig. 5 (b), the complexity of the F-BB scheme becomes prohibitive and the number of required iterations cannot be shown. Compared to that, it takes up to only 12 iterations for the P-BB method to converge. 

\begin{figure}[!t]
\centering
\includegraphics[scale=0.4]{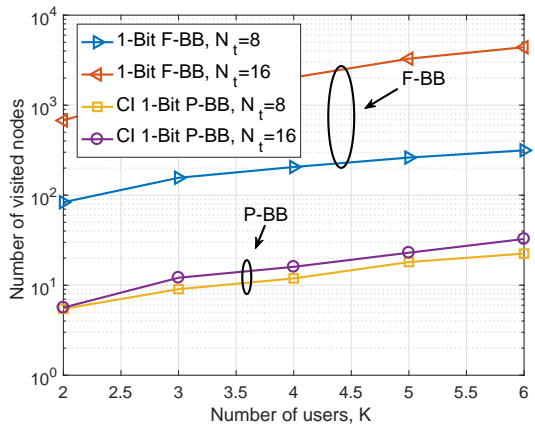}
\caption{Number of visited nodes v.s. number of users $K$, QPSK, $N_t=8$ and $N_t=16$}
\end{figure} 

To further reveal the complexity gain of the proposed P-BB algorithm, we compare the total number of visited nodes for P-BB and F-BB methods with respect to the increasing number of users in Fig. 6, where we consider two scenarios with $N_t=8$ and $N_t=16$, respectively. For both cases, it is apparent that the number of visited nodes for P-BB is much fewer than that for F-BB, especially when $N_t$ becomes larger. Both the results in Fig. 5 and Fig. 6 demonstrate the significant complexity reduction of the proposed P-BB approach. 

\begin{figure}[!t]
\centering
\includegraphics[scale=0.4]{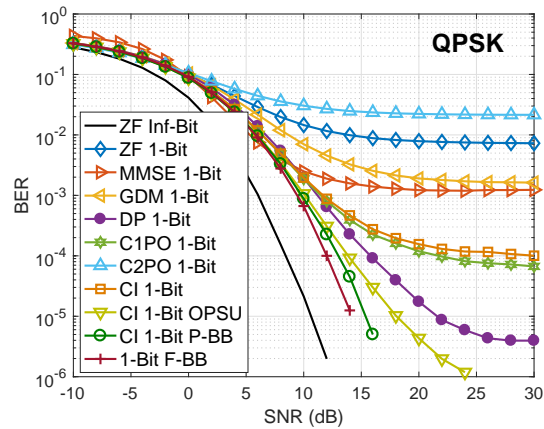}
\caption{BER v.s. transmit SNR, $N_t=8$, $K=2$, QPSK, $\text{N}_\text{DP}=50$, $\text{N}_\text{C1PO}=25$, $\text{N}_\text{C2PO}=25$}
\end{figure} 

For QPSK modulation, we depict the BER result of a small-scale MIMO system in Fig. 7 for $K=2$ and $N_t=8$, where we have also included the optimal F-BB method for comparison. A general observation is that quantized non-linear precoding methods perform better than quantized linear precoding methods, especially when the transmit SNR exceeds 10dB. Among the non-linear precoding schemes, it is observed that the proposed `CI 1-Bit P-BB' achieves the best BER performance, with only less than 1dB SNR loss compared to the optimal F-BB method `1-Bit F-BB', which demonstrates its superiority. For the proposed low-complexity `CI 1-Bit OPSU', while it achieves a slightly inferior performance to the proposed P-BB method, it is more computationally efficient by avoiding the BB process. Compared to the `DP 1-Bit' method which performs the best among the existing 1-bit precoding approaches for a small-scale MIMO system, it is worth highlighting that the SNR gain of the 1-bit precoding methods proposed in this paper can be as large as 5dB when the BER is $10^{-4}$, and becomes more prominent when the BER goes lower.

\begin{figure}[!t]
\centering
\includegraphics[scale=0.4]{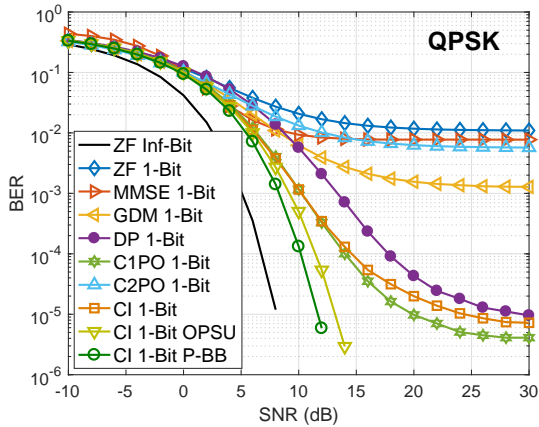}
\caption{BER v.s. transmit SNR, $N_t=64$, $K=16$, QPSK, $\text{N}_\text{DP}=50$, $\text{N}_\text{C1PO}=25$, $\text{N}_\text{C2PO}=25$}
\end{figure} 

We extend our BER result for QPSK modulation to the case of massive MIMO in Fig. 8, where a $64 \times 16$ MIMO system is considered. In such scenarios, the optimal F-BB method is no longer applicable due to its prohibitive computational cost. When the system scales up, we observe that the performance gains of the 1-bit precoding methods proposed in this paper become larger compared to the existing works. More specifically, compared to `C1PO 1-Bit' which achieves the best BER performance among the existing works in this scenario, we observe an SNR gain up to more than 7dB for the proposed methods in this paper, when the BER is $10^{-5}$.

In what follows, we consider massive MIMO systems with higher-order PSK modulations, where existing 1-bit precoding methods usually exhibit poor error rate performance. In Fig. 9, we depict the BER result for a $128 \times 16$ MIMO system when 8PSK modulation is employed. In this scenario, among the existing works we observe that only `CI 1-Bit' and `C1PO 1-Bit' can achieve acceptable error rate performance. Similar to the case when QPSK modulation is adopted, both 1-bit precoding methods proposed in this paper exhibit superior BER performance, with an SNR gain up to more than 5dB compared to `CI 1-Bit'.

\begin{figure}[!t]
\centering
\includegraphics[scale=0.4]{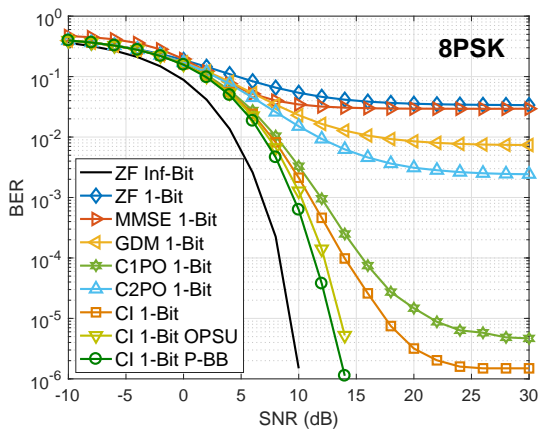}
\caption{BER v.s. transmit SNR, $N_t=128$, $K=16$, 8PSK, $\text{N}_\text{C1PO}=25$, $\text{N}_\text{C2PO}=25$}
\end{figure} 

\begin{figure}[!t]
\centering
\includegraphics[scale=0.4]{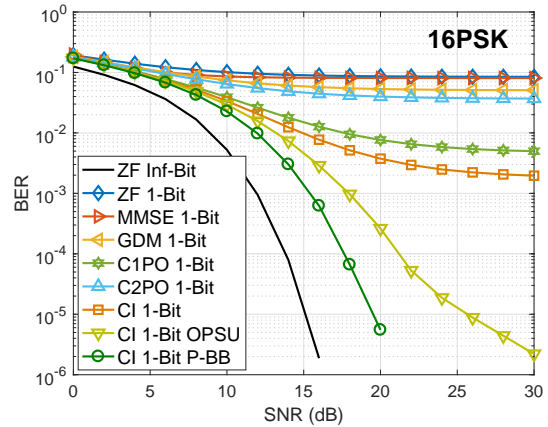}
\caption{BER v.s. transmit SNR, $N_t=128$, $K=16$, 16PSK, $\text{N}_\text{C1PO}=25$, $\text{N}_\text{C2PO}=25$}
\end{figure} 

Fig. 10 further depicts the error rate performance for 16PSK modulation in a $128 \times 16$ MIMO system. In this scenario, we observe an error floor for all the existing 1-bit precoding methods, and the best BER they achieve cannot be lower than $10^{-3}$. As a comparison, both 1-bit precoding schemes proposed in this paper exhibit promising BER results, achieving a BER lower than $10^{-5}$ when the transmit SNR is equal to 20dB and 25dB, respectively. Moreover, as opposed to results for QPSK and 8PSK where the proposed methods achieve comparable BER results, we observe that the performance gap between the proposed `CI 1-Bit P-BB' and the proposed `CI 1-Bit OPSU' becomes more significant, when the higher-order 16PSK modulation is adopted.

\subsection{Results for QAM}

\begin{figure}[!t]
\centering
\subfloat[$N_t=8$, $K=2$]
{%
  \includegraphics[clip,width=0.9\columnwidth]{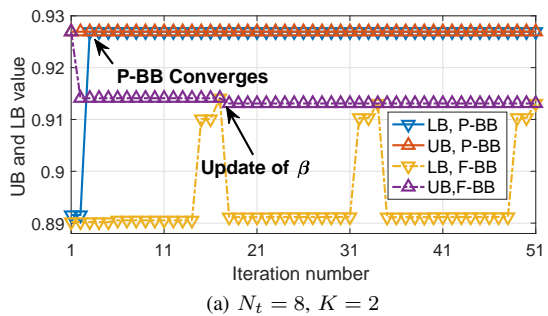}%
}

\subfloat[$N_t=64$, $K=8$]
{%
  \includegraphics[clip,width=0.9\columnwidth]{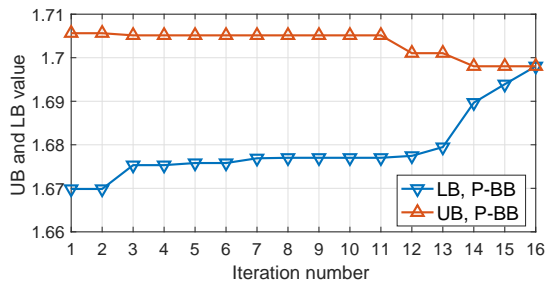}%
}
\caption{Convergence of the P-BB method, 16QAM, SNR=0dB, $\epsilon_0=10^{-3}$}
\end{figure}

\begin{figure}[!t]
\centering
\includegraphics[scale=0.4]{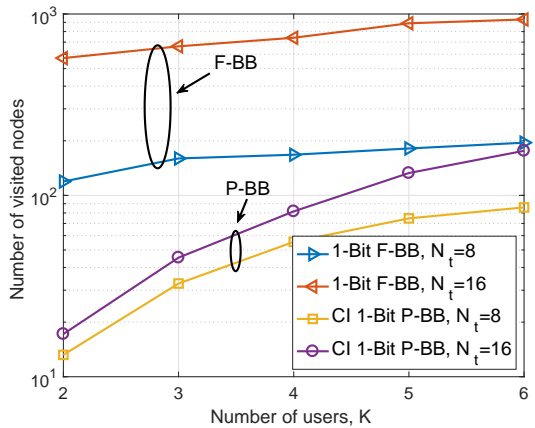}
\caption{Number of visited nodes v.s. number of users $K$, 16QAM, $N_t=8$ and $N_t=16$, SNR=0dB, $\epsilon_0=10^{-3}$}
\end{figure} 

We move to the description for the numerical results when QAM modulation is considered at the BS. Similar to the case of PSK, we first demonstrate the convergence of the proposed P-BB approach before presenting the BER results, as shown in Fig. 11 for 16QAM when the transmit SNR is 0dB, where we note that the `iteration number' is the total number of iterations including both the alternating optimization and the P-BB process. In Fig. 11 (a) where $N_t=8$ and $K=2$, i.e., a small-scale MIMO case, we observe that the P-BB method becomes convergent within 4 iterations, while the F-BB approach requires more than 50 iterations to become convergent. When the considered scenario scales up to a $64\times8$ MIMO system where the F-BB scheme becomes inapplicable, we observe in Fig. 11 (b) that the required number of iterations for convergence of the proposed P-BB method is only 16.  

\begin{figure}[!t]
\centering
\includegraphics[scale=0.4]{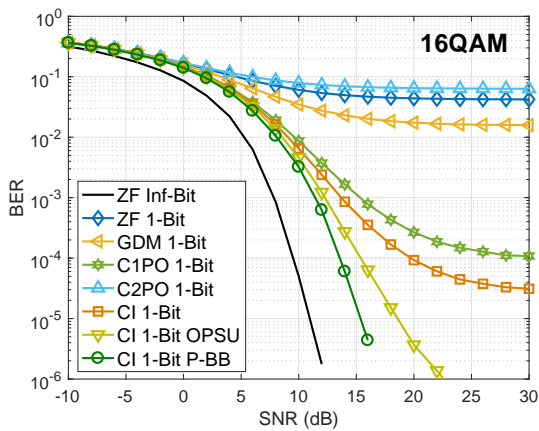}
\caption{BER v.s. transmit SNR, $N_t=64$, $K=8$, 16QAM, $\text{N}_\text{C1PO}=25$, $\text{N}_\text{C2PO}=25$, $\epsilon_0=10^{-3}$}
\end{figure} 

To demonstrate the complexity gain of P-BB compared to F-BB for QAM signaling, Fig. 12 further depicts the number of visited nodes with respect to the increasing number of users, where the cases of both $N_t=8$ and $N_t=16$ are considered for 16QAM modulation. Similar to the result when PSK signaling is considered, as shown in Fig. 6, we observe a considerable complexity reduction for the proposed P-BB process, especially when the number of users is small.

In Fig. 13, we present the BER result for a $64 \times 8$ MIMO system when 16QAM modulation is adopted at the BS. Similar to the case of PSK, we observe significant performance improvements for both 1-bit precoding schemes proposed in this paper compared to `CI 1-Bit' and `C1PO 1-Bit' in the literature, where the SNR gain can be as large as 5dB when the BER is below $10^{-4}$, which demonstrates the superiority of the proposed schemes for QAM modulation.

\begin{figure}[!t]
\centering
\includegraphics[scale=0.4]{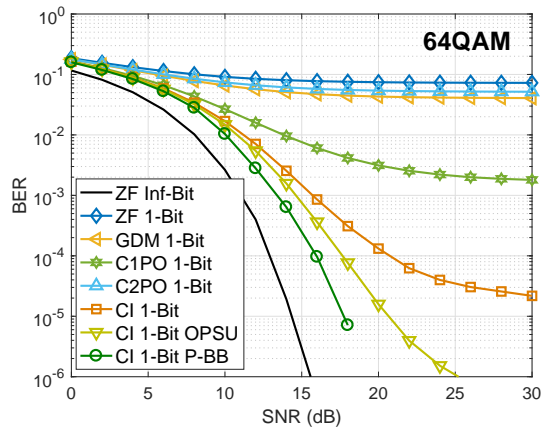}
\caption{BER v.s. transmit SNR, $N_t=128$, $K=8$, 64QAM, $\text{N}_\text{C1PO}=25$, $\text{N}_\text{C2PO}=25$, $\epsilon_0=10^{-3}$}
\end{figure} 

Fig. 14 further depicts the BER result when 64QAM is employed for a $128 \times 8$ MIMO system, where error floors are observed for existing 1-bit precoding approaches. Similar to the result for 16QAM, we observe substantial error rate improvements for the proposed 1-bit precoding methods based on P-BB and OPSU, compared to existing linear 1-bit approaches and the conventional `CI 1-Bit' scheme. In particular, the SNR gain can be more than 5dB when the BER is lower than $10^{-4}$.

\section{Conclusion}
In this paper, we have proposed several 1-bit precoding approaches for massive MIMO downlink based on CI, where both PSK and QAM signaling are considered. The proposed 1-bit precoding methods are based on the observation that most entries in the output signals obtained from solving the relaxed CI-based optimization problem already satisfy the 1-bit requirement. Therefore, the BB operation as well as the sequential update operation are only applied to a small portion of the entries of the output signals that do not comply with the 1-bit requirement, which leads to significant savings in terms of computational complexity. Simulation results have validated the effectiveness of the proposed 1-bit precoding algorithms, which demonstrate superior error rate performance.

\appendices
\section{Proof for Lemma 1}
For notational simplicity, we first introduce 
\begin{equation}
\alpha_0={{\Re \left( {s_k^{\cal A}} \right)\Im \left( {s_k^{\cal B}} \right) - \Im \left( {s_k^{\cal A}} \right)\Re \left( {s_k^{\cal B}} \right)}}.
\label{eq_16}
\end{equation}
Subsequently, we can further transform the expressions for $\alpha_k^{\cal A}$ and $\alpha_k^{\cal B}$ in \eqref{eq_11} into
\begin{equation}
\begin{aligned}
\alpha _k^{\cal A}&=\left[ {\frac{{\Im \left( {s_k^{\cal B}} \right)}}{{{\alpha _0}}}, - \frac{{\Re \left( {s_k^{\cal B}} \right)}}{{{\alpha _0}}}} \right]\left[ {\begin{array}{*{20}{c}}
{\Re \left( {{\bf{h}}_k^\text{T}} \right)}&{ - \Im \left( {{\bf{h}}_k^\text{T}} \right)}\\
{\Im \left( {{\bf{h}}_k^\text{T}} \right)}&{\Re \left( {{\bf{h}}_k^\text{T}} \right)}
\end{array}} \right] {\bf x_E} \\
&= {\bf{u}}_k^\text{T}{\bf{H}}_{k,{\text E}}^\text{T}{{\bf{x}}_{\bf{E}}},
\end{aligned}
\label{eq_17}
\end{equation}
and 
\begin{equation}
\begin{aligned}
\alpha _k^{\cal B}&= \left[ { - \frac{{\Im \left( {s_k^{\cal A}} \right)}}{{{\alpha _0}}},\frac{{\Re \left( {s_k^{\cal A}} \right)}}{{{\alpha _0}}}} \right]\left[ {\begin{array}{*{20}{c}}
{\Re \left( {{\bf{h}}_k^\text{T}} \right)}&{ - \Im \left( {{\bf{h}}_k^\text{T}} \right)}\\
{\Im \left( {{\bf{h}}_k^\text{T}} \right)}&{\Re \left( {{\bf{h}}_k^\text{T}} \right)}
\end{array}} \right] {\bf x_E} \\
&={\bf{v}}_k^\text{T}{\bf{H}}_{k,{\text E}}^\text{T}{{\bf{x}}_{\bf{E}}},
\end{aligned}
\label{eq_18}
\end{equation}
where ${\bf H}_{k, {\text E}}^\text{T}$ expands ${\bf h}_k^\text{T}$ into its real equivalence, ${\bf u}_k$ and ${\bf v}_k$ are given by
\begin{equation}
{{\bf{u}}_k} = {\left[ {\frac{{\Im \left( {s_k^{\cal B}} \right)}}{{{\alpha _0}}}, - \frac{{\Re \left( {s_k^{\cal B}} \right)}}{{{\alpha _0}}}} \right]^\text{T}}, {\kern 3pt} {{\bf{v}}_k} = {\left[ { - \frac{{\Im \left( {s_k^{\cal A}} \right)}}{{{\alpha _0}}},\frac{{\Re \left( {s_k^{\cal A}} \right)}}{{{\alpha _0}}}} \right]^\text{T}}.
\label{eq_19}
\end{equation} 
Based on this transformation, we can express $\bf M$ as
\begin{equation}
{\bf M}=\left[ {\begin{array}{*{20}{c}}
{{\bf{u}}_1^\text{T}}&{\bf{0}}& \cdots &{\bf{0}}\\
{\bf{0}}&{{\bf{u}}_2^\text{T}}& \ddots & \vdots \\
 \vdots & \ddots & \ddots &{\bf{0}}\\
{\bf{0}}& \cdots &{\bf{0}}&{{\bf{u}}_K^\text{T}}\\
{{\bf{v}}_1^\text{T}}&{\bf{0}}& \cdots &{\bf{0}}\\
{\bf{0}}&{{\bf{v}}_2^\text{T}}& \ddots & \vdots \\
 \vdots & \ddots & \ddots &{\bf{0}}\\
{\bf{0}}& \cdots &{\bf{0}}&{{\bf{v}}_K^\text{T}}
\end{array}} \right]\left[ {\begin{array}{*{20}{c}}
{{\bf{H}}_{1,{\text E}}^\text{T}}\\
{{\bf{H}}_{2,{\text E}}^\text{T}}\\
 \vdots \\
{{\bf{H}}_{K,{\text E}}^\text{T}}
\end{array}} \right]= {\bf{T G}},
\label{eq_20}
\end{equation}
with ${\bf T} \in {\cal R}^{2K \times 2K}$ and ${\bf G} \in {\cal R}^{2K \times 2N_t}$. Based on the expressions for ${\bf u}_k$ and ${\bf v}_k$ in \eqref{eq_19}, we observe that there does not exist a constant $c$ such that ${\bf v}_k = c \cdot {\bf u}_k$, $\forall k$, since ${\bf v}_k$ and ${\bf u}_k$ are the two bases for $s_k$ as in \eqref{eq_5} and are not parallel. Therefore, ${\bf T}$ is full-rank, i.e., ${\text {rank}}\left( {\bf T} \right)=2K$, which further means that ${\text {rank}}\left( {\bf M} \right)={\text {rank}}\left( {\bf G} \right)$ \cite{book1}. We further obtain ${\text {rank}}\left( {\bf G} \right)={\text {rank}}\left( {\bf H_E} \right)$, where 
\begin{equation}
{{\bf{H}}_{\bf{E}}} = \left[ {\begin{array}{*{20}{c}}
{\Re \left( {\bf{H}} \right)}&{ - \Im \left( {\bf{H}} \right)}\\
{\Im \left( {\bf{H}} \right)}&{\Re \left( {\bf{H}} \right)}
\end{array}} \right],
\label{eq_21}
\end{equation}
which expands $\bf H$ into its real equivalence, based on the observation that $\bf G$ is identical to $\bf H_E$ after some row rearrangements. Since the flat-fading Rayleigh fading channel $\bf H$ is full-rank with probability 1, and the real part and imaginary part of $\bf H$ are i.i.d., we obtain the rank of $\bf M$ is ${\text {rank}}\left( {\bf H_E} \right)=2 \cdot {\text {rank}}\left( {\bf H} \right) =2K$ with probability 1, which completes the proof. ${\kern 215pt} \blacksquare$

\section{Proof for Proposition 1}
Proving this proposition is equivalent to proving that the number of entries whose amplitudes are strictly smaller than $\frac{1}{\sqrt {2N_t}}$ is at most $\left( {2K - 1} \right)$.

Firstly, we transform the convex optimization problem ${\cal P}_3$ into a standard minimization form, given by
\begin{equation}
\begin{aligned}
&\mathcal{P}_{12}: {\kern 3pt} \mathop {\min }\limits_{{{{\bf{\tilde x}}}_{\bf{E}}}} {\kern 3pt} -t \\
&{\kern 5pt} \text{s.t.} {\kern 10pt} t - {\bf{m}}_l^\text{T}{{\bf \tilde x}_{\bf{E}}} \le 0, {\kern 3pt} \forall l \in {\cal L} \\
&{\kern 27pt} \tilde x_m^{\text E} - \frac{1}{{\sqrt {2{N_t}} }} \le 0, {\kern 3pt} \forall m \in {\cal M} \\
&{\kern 25pt} -\tilde x_m^{\text E} - \frac{1}{{\sqrt {2{N_t}} }} \le 0, {\kern 3pt} \forall m \in {\cal M}
\label{eq_23}
\end{aligned}
\end{equation}
The corresponding Lagrangian of ${\cal P}_{12}$ can be constructed as
\begin{equation}
\begin{aligned}
&{\cal L}\left( {t,{{{\bf{\tilde x}}}_{\bf{E}}},{\beta _l},{\mu _m},{\nu _m}} \right) =  - t + \sum\limits_{l = 1}^{2K} {{\beta _l}\left( {t - {\bf{m}}_l^\text{T}{{\bf \tilde {x}}_{\bf{E}}}} \right)} \\
& {\kern 30pt} + \sum\limits_{m = 1}^{2{N_t}} {{\mu _m}\left( {\tilde x_m^{\text E} - \frac{1}{{\sqrt {2{N_t}} }}} \right)}  - \sum\limits_{m = 1}^{2{N_t}} {{\nu _m}\left( {\tilde x_m^{\text E} + \frac{1}{{\sqrt {2{N_t}} }}} \right)}\\
&=\left( {{{\bf{1}}^\text{T}}{\bm \beta}  - 1} \right)t - {{\bm \beta} ^\text{T}}{\bf{M}}{{\bf{\tilde x}}_{\bf{E}}} + \left( {{{\bm \mu} ^\text{T}} - {{\bm \nu} ^\text{T}}} \right){{\bf{\tilde x}}_{\bf{E}}} \\
& {\kern 12pt} - \frac{1}{{\sqrt {2{N_t}} }}\left( {{{\bf{1}}^\text{T}}{\bm \mu}  + {{\bf{1}}^\text{T}}{\bm \nu} } \right),
\end{aligned}
\label{eq_24}
\end{equation}
where ${\bm \beta} \in {\cal R}^{2K \times 1}$, ${\bm \mu} \in {\cal R}^{2N_t \times 1}$, and ${\bm \nu} \in {\cal R}^{2N_t \times 1}$. Accordingly, we formulate the KKT conditions as
\begin{IEEEeqnarray}{rCl} 
\IEEEyesnumber
\frac{{\partial {\cal L}}}{{\partial t}} =  {{\bf{1}}^\text{T}}{\bm \beta}  - 1  = 0 {\kern 30pt} \IEEEyessubnumber* \label{eq_25a} \\
\frac{{\partial {\cal L}}}{{\partial {{{\bf{\tilde x}}}_{\bf{E}}}}} =  - {{\bf{M}}^\text{T}}{\bm \beta}  + {\bm \mu}  - {\bm \nu}  = {\bf{0}} {\kern 30pt} \label{eq_25b} \\
{{\beta _l}\left( {t - {\bf{m}}_l^\text{T}{{\bf \tilde {x}}_{\bf{E}}}} \right)}=0, {\kern 3pt} \beta_l \ge 0, {\kern 3pt} \forall l \in {\cal L} {\kern 30pt} \label{eq_25c} \\
{{\mu _m}\left( {\tilde x_m^{\text E} - \frac{1}{{\sqrt {2{N_t}} }}} \right)}=0, {\kern 3pt} {\mu _m}\ge 0, {\kern 3pt} \forall m \in {\cal M} {\kern 30pt} \label{eq_25d} \\
{{\nu _m}\left( {\tilde x_m^{\text E} + \frac{1}{{\sqrt {2{N_t}} }}} \right)}=0, {\kern 3pt} {\nu _m}\ge 0, {\kern 3pt} \forall m \in {\cal M} {\kern 30pt} \label{eq_25e}
\end{IEEEeqnarray}

In the following, we prove this proposition by contradiction. Suppose that there are a total number of $2K$ entries in ${\bf \tilde x}_{\bf E}$ whose amplitudes are strictly smaller than $\frac{1}{\sqrt {2N_t}}$. For notational simplicity, we introduce a set $\cal S$ that consists of the indices of these entries in ${\bf \tilde x}_{\bf E}$, which can be mathematically expressed as
\begin{equation}
n \in {\cal S}, {\kern 3pt} {\text {if}} {\kern 3pt} \left| {\tilde x_n^{\text E}} \right| < \frac{1}{{\sqrt {2{N_t}} }},
\label{eq_26}
\end{equation}
and based on our above assumption we have $\text{card}\left( {\cal S} \right)=2K$. According to the complementary slackness conditions \eqref{eq_25d} and \eqref{eq_25e}, we further obtain
\begin{equation}
\mu _n=0, {\kern 3pt} \nu_n=0, {\kern 3pt} \forall n \in {\cal S}.
\label{eq_27}
\end{equation}
Recall \eqref{eq_25b} which can be viewed as a system of linear equations with $\bm \beta$ being the variable, and for simplicity we introduce ${\bf W}={\bf M}^\text{T}={\left[ {{\bf{w}}_1,{\bf{w}}_2, \cdots ,{\bf{w}}_{2{N_t}}} \right]^\text{T}}$. We subsequently pick the corresponding rows of ${\bf W}$ whose indices belong to $\cal S$ to formulate a subsystem of linear equations, given by
\begin{equation}
{\bf W_p}{\bm \beta}  = \hat {\bm \mu}_{\bf p}  - \hat {\bm \nu}_{\bf p}  = {\bf{0}},
\label{eq_28}
\end{equation}
where ${\bf W_p} \in {\cal R}^{\text{card}\left( {\cal S} \right) \times 2K}$ is expressed as
\begin{equation}
{\bf W_p} = {\left[ {{\bf{w}}_{{n_1}}, \cdots ,{\bf{w}}_{{n_m}}, \cdots ,{\bf{w}}_{{n_{\text{card}\left( {\cal S} \right)}}}} \right]^\text{T}}, {\kern 3pt} \forall n_m \in {\cal S}.
\label{eq_29}
\end{equation}
Based on the result in {\bf Lemma 1} and that $\text{card}\left( {\cal S}\right) = 2K$, we obtain that ${\bf W_p}$ is full-rank. According to the linear algebra theory \cite{book1}, given a full-rank coefficient matrix $\bf W_p$, a non-zero solution to \eqref{eq_28} does not exist and there is only a trivial solution, i.e.,
\begin{equation}
{\bm \beta}^*={\bf 0}.
\label{eq_30}
\end{equation}
However, this solution does not comply with \eqref{eq_25a} that enforces a non-zero solution of $\bm \beta$, which causes contradiction. By following a step similar to the above, this contradiction is also observed if we assume there are a total number of $N>2K$ entries in the obtained $\bf \tilde x_E$ whose amplitudes are strictly smaller than $\frac{1}{\sqrt{2N_t}}$, which completes the proof. {\kern 11pt} $\blacksquare$

\ifCLASSOPTIONcaptionsoff
  \newpage
\fi

\bibliographystyle{IEEEtran}
\bibliography{refs.bib}

\end{document}